\documentclass{jpp}

\usepackage{amsmath}
\usepackage{amssymb}
\usepackage{graphicx}
\usepackage{bm}
\usepackage{mathrsfs}
\usepackage{color}
\usepackage{subfigure}
\usepackage{epstopdf, epsfig}
\usepackage{mathabx}
\usepackage{array} 
\usepackage{natbib}

\newcommand{\beq}{\begin{equation}}
\newcommand{\eeq}{\end{equation}}
\newcommand{\wbpar}{\widetilde{B}_\parallel}
\newcommand{\bpar}{B_\parallel}
\newcommand{\apar}{ A_{\parallel}}
\newcommand{\wapar}{ \widetilde{A}_{\parallel}}
\newcommand{\wphi}{\widetilde{\phi}}

\newcommand{\lapp}{\nabla_{\perp}^2}

\newcommand{\pa}{\partial}
\newcommand{\dfa}{\widetilde{f}_{s}}
\newcommand{\qa}{q_{s}}

\newcommand{\vtpa}{v_{{th }_{\parallel s}}}
\newcommand{\Tpea}{T_{{0 }_{\perp s}}}
\newcommand{\Tpa}{T_{{0 }_{\parallel s}}}

\newcommand{\Tpee}{T_{{0 }_{\perp e}}}
\newcommand{\Tpae}{T_{{0 }_{\parallel e}}}

\newcommand{\taupa}{\tau_{\perp_s}}
\newcommand{\taupi}{\tau_{\perp_i}}
\newcommand{\calfa}{\mathcal{F}_{{eq}_s}}

\newcommand{\vpar}{v_\|}
\newcommand{\aal}{a_{s}}
\newcommand{\dwa}{d \mathcal{W}_s}
\newcommand{\thea}{\Theta_s}

\newcommand{\bepea}{\beta_{\perp_s}}
\newcommand{\nno}{\nonumber}
\newcommand{\cspe}{c_{s_\perp}}

\newcommand{\gamu}{G_{10}}
\newcommand{\gamd}{G_{20}}
\newcommand{\gamz}{\Gamma_0}
\newcommand{\gammu}{\Gamma_1}
\newcommand{\ba}{b_s}
\newcommand{\bepe}{\beta_{\perp_e}}

\newcommand{\sgn}{\mathrm{sgn}}
\newcommand{\bepa}{\beta_{\perp_s}}

\newcommand{\mua}{\mu_s}
\newcommand{\ga}{\widetilde{g}_s}
\newcommand{\hz}{\hat{z}}

\newcommand{\bk}{\boldsymbol{k}}

\newcommand{\bx}{\boldsymbol{x}}
\newcommand{\rsth}{\rho_{\perp_s}}
\newcommand{\ben}{\begin{eqnarray}}
\newcommand{\een}{\end{eqnarray}}

\newcommand{\jo}{J_0(\aal)}
\newcommand{\ju}{J_1(\aal)}

\newcommand{\bbk}{\boldsymbol{k}}
\newcommand{\phik}{\wphi_{\boldsymbol{k}}}
\newcommand{\ak}{\widetilde{A}_{\parallel \bbk}}
\newcommand{\bpark}{\widetilde{B}_{\parallel \bbk}}

\newcommand{\rthpes}{\rho_{th_{ \perp s}}}

\newcommand{\rs}{\rho_s}

\newcommand{\Tpi}{T_{\parallel_i}}
\newcommand{\Tpe}{T_{\parallel_e}}

\newcommand{\cald}{\mathcal{D}}
\newcommand{\hatz}{\hat{z}}
\newcommand{\wns}{\widetilde{N}_s}
\newcommand{\wus}{\widetilde{U}_s}
\newcommand{\wtps}{\widetilde{T}_{\parallel s}}
\newcommand{\wtpes}{\widetilde{T}_{\perp s}}
\newcommand{\Tps}{T_{\parallel s}}

\newcommand{\rthpei}{\rho_{th_{\perp_i}}}
\newcommand{\dfk}{\widetilde{f}_{s_{\bk}}}
\newcommand{\scrd}{\mathscr{D}}
\newcommand{\zs}{\widetilde{\zeta_s}}
\newcommand{\fks}{f_{{mn}_{s \bk}}}
\newcommand{\sgq}{\sgn(\qa)}
\newcommand{\sumz}{\sum_{n=0}^{+\infty}}
\newcommand{\qpars}{\widetilde{Q}_{\parallel s}}
\newcommand{\qps}{Q_{\parallel s}}

\newcommand{\gamus}{G_{10s}}

\newcommand{\gamui}{G_{10i}}
\newcommand{\gamds}{G_{20s}}

\newcommand{\gamzs}{\Gamma_{0s}}
\newcommand{\gammus}{\Gamma_{1s}}

\newcommand{\us}{\mathbf{u}_{\perp_s}}
\newcommand{\epars}{E_{\parallel s}}
\newcommand{\gpars}{\nabla_{\parallel s}}

	\title{A Hamiltonian gyrofluid model based on a quasi-static closure}

	\author{E. Tassi$^1$, T. Passot$^1$ and P.L. Sulem$^1$}

	\affiliation{$^1$ Universit\'e C\^ote d'Azur, CNRS, Observatoire de la C\^ote d'Azur, Laboratoire J.L. Lagrange, Boulevard de l'Observatoire, CS  34229, 06304 Nice Cedex 4, France}
	
\begin{document}
		
		\maketitle
		
\begin{abstract}

A Hamiltonian six-field gyrofluid model is constructed, based on closure relations derived from the so-called "quasi-static" gyrokinetic linear theory
where the fields are assumed to propagate with a parallel phase velocity much smaller than the parallel particle thermal velocities. The main properties captured by this model, primarily aimed at exploring fundamental problems of interest for space plasmas such as the solar wind, are its ability to provide a reasonable agreement with kinetic theory for linear low-frequency modes, and at the same time to ensure a Hamiltonian structure in the absence of explicit dissipation. The model accounts for equilibrium temperature anisotropy, ion and electron finite Larmor radius corrections, electron inertia, magnetic fluctuations along the direction of a strong guide field, and parallel Landau damping, introduced through a Landau-fluid modeling of the parallel heat transfers for both gyrocenter species. Remarkably, the quasi-static closure leads to exact and simple expressions for the nonlinear terms involving gyroaveraged electromagnetic fields and potentials. One of the consequences is that a rather natural identification of the Hamiltonian structure of the model becomes possible when Landau damping is neglected. A slight variant of the model consists of a four-field Hamiltonian reduction of the original six-field model, which is also used for the subsequent linear analysis. In the latter, the dispersion relations of kinetic Alfv\'en waves and the firehose instability are shown to be correctly reproduced, relatively far in the sub-ion range (depending on the plasma parameters), while the spectral range where the slow-wave dispersion relation and the field-swelling instabilities are precisely described is less extended.  This loss of accuracy originates from the breaking of the condition of small phase velocity, relative to the parallel thermal velocity of the electrons (for kinetic Alfv\'en waves and firehose instability) or of the ions (in the case of the field-swelling instabilities).

\end{abstract}

\section{Introduction}
Modeling the dynamics of collisionless (or weakly collisional) plasmas at scales comparable to or smaller than the ion Larmor radius is an important issue both for laboratory and astrophysical plasmas. At the level of a kinetic description, a valuable tool is given by the gyrokinetic theory which provides a reduction of the Vlasov-Maxwell (VM) equations by focusing on phenomena with a characteristic time scale large compared with the ion gyro-period. This approach, which eliminates the dependency on the gyration angle, typically adopts,  as dynamical variables, the distribution functions of the gyrocenters rather than those of the particles. Within the gyrokinetic framework, a subset of models consists of the so-called $\delta f$-gyrokinetic models, which assume the gyrocenter distribution functions of the various particle species, to be close to those of an equilibrium state. In spite of this reduction, numerical simulations of three-dimensional gyrokinetic equations  in a turbulent regime (even in the $\delta f$ framework) require huge computational resources, which justifies the development of simpler (although less complete) descriptions based on gyrofluid equations governing the evolution of a finite number of  moments of the gyrocenter distribution functions. The relation between gyrocenter and particle moments is well-defined and can usually  be computed perturbatively. As in the case of the fluid hierarchy derived from the VM equations, a gyrofluid hierarchy of equations needs to be closed, in order to obtain a gyrofluid model with a finite number of dynamical variables. An important condition  to prescribe at the level of closure assumptions is the preservation, in the absence of dissipation, of the Hamiltonian character of the parent gyrokinetic equations.  This guarantees that in the reduction from a gyrokinetic to a gyrofluid system, not only no uncontrolled dissipation of the total energy is introduced but also that further invariants (Casimir invariants) of the system exist, and that the dynamics takes place on hyper-surfaces in phase space where the values of these invariants are constant. Another constraint is the consistency with  the linear gyrokinetic theory. In particular, this requires retaining the influence of resonant effects such as Landau damping. A closure accounting for Landau damping in the $\delta f$ approach typically introduces dissipation and thus prevents the model from being Hamiltonian. Such form of dissipation is, however, voluntarily added and the main requirement is that the model possesses a Hamiltonian structure when the dissipative terms are removed.

In this spirit, the main goal of the present paper is to construct a gyrofluid model possessing the above mentioned properties, and primarily addressed to study phenomena relevant for collisionless space plasmas. Motivated by  measurements of sub-proton fluctuations in the solar wind (see \citet{Sah10} and \citet{Alexandrova13} or \citet{Bruno13} for reviews),
reduced fluid models have already been derived and numerically integrated to explore the dynamics of space plasmas. Kinetic Alfv\'en wave (KAW) turbulence was for example addressed in \citet{Boldyrev-Perez12}. A more general Hamiltonian reduced gyrofluid model \citep{PST18} which, in the appropriate asymptotic limit yields the model of \citet{Boldyrev-Perez12}, was recently developed to simultaneously capture the three regimes of: dispersive Alfv\'en waves (at scales larger than the sonic and/or ion Larmor radius), of KAWs at sub-ion scales, and also of inertial kinetic Alfv\'en waves  (at scales comparable to the electron inertial length \citep{Chen-Bold17,PST17}). It has  been used to derive weak turbulence kinetic equations \citep{PS19} and the properties of imbalanced KAW turbulence in the framework of a reduction to nonlinear diffusion equations in spectral space \citep{MPS20}. Nevertheless, to the best of our knowledge, existing reduced gyrofluid models  can capture parallel magnetic field fluctuations, electron inertia and ion finite Larmor radius effects, but (with the exception of the recent models of \citet{Tas19}, which will be discussed in Sec. \ref{sssec:gf4}) do not take into account a possible temperature anisotropy of the equilibrium state.  It however turns out that collisionless space plasmas, such as the solar wind, often exhibit anisotropic distribution functions \citep{Marsch12} that can result from various heating effects such as Landau damping \citep{Chen19} or stochastic heating \citep{Bourouaine13, Hoppock18} or from mechanical effects such as the action of a shear flow \citep{DeCamillis16}. Proton temperature anisotropies play an important role in the solar wind \citep{Hellinger06}, and possibly even more so at closer distance from the Sun \citep{Huang2020}. Temperature anisotropies are usually constrained by the micro-instabilities they trigger (e.g. mirror and firehose instabilities), both for ions \citep{Bale09} and electrons \citep{Stverak08}. The range of these accessible temperature anisotropies increases as the beta parameter (ratio of thermal to magnetic presure) decreases. In addition, temperature anisotropies are known to affect the development of the tearing instability \citep{Shi87}, and thus the stability of current sheets \citep{Matteini13}, which plays a major role in the turbulence evolution \citep{Franci17}. Another important feature to account for is the coupling to ion acoustic waves, which permits the development of the parametric decay of Alfv\'en waves at the MHD scales at small beta
\citep{DelZanna01}, an effect that  contributes to the generation of counter propagating waves required for the development of a turbulent dynamics at these scales.
Interestingly, this parametric instability, proposed by \citet{Bowen18} for the generation of the solar wind compressive fluctuations, turns out to occur in a wider range of beta parameters in the presence of temperature anisotropy \citep{Tenerani17}. A useful feature of a new gyrofluid model for space plasma studies would thus be its capability to account for equilibrium temperature anisotropies and the coupling to ion acoustic waves. 
	
As starting point for the derivation of the model, we choose the $\delta f$ gyrokinetic equations presented in \citet{Kun15} where, for the sake of simplicity, we assume an electron-proton plasma with an equilibrium state described by bi-Maxwellian distribution functions with no mean drift velocity. Such a system fully satisfies our requirements. Indeed, it is a $\delta f$ gyrokinetic model mainly conceived for pressure-anisotropic astrophysical plasmas and, as such, it specifically accounts for equilibrium temperature anisotropy (unlike most of gyrokinetic models which consider a generic equilibrium distribution function or specialize to the case of a Maxwellian equilibrium). Also, it accounts for parallel magnetic perturbations and it has been shown to possess (at least in the limit of interest for our derivation)  a Hamiltonian structure \citep{Tas19}. As far as  the number of moments to be retained in the gyrofluid model is concerned, the inclusion of Landau damping requires to retain at least the first three moments for each particle species. This is why, under the assumption of a two-species plasma, we opt for the derivation of a six-field gyrofluid   model  evolving three moments, including parallel temperature fluctuations, for each species. Nevertheless, a four-field Hamiltonian reduced version will also be presented and applied. Another novelty with respect to already existing Hamiltonian gyrofluid models, is the adoption of a closure relation, referred to as {\it quasi-static}, derived from linear gyrokinetic theory in the limit of slowly-evolving fields. More precisely, according to such closure, all gyrofluid moments that are not determined by gyrofluid evolution equations, are fixed according to their expression obtained from the gyrokinetic linear theory in the limit $|\omega/(k_z \vtpa)| \ll1$, where $\omega$ is the frequency of a mode, $k_z$ the component of its wave vector along the direction of the guide field, and $\vtpa$ is the thermal speed of the species $s$, associated with the equilibrium temperature along the direction of the guide field. As such, this closure is suitable for fields slowly-evolving (i.e. quasi-static) with respect to particles travelling at the parallel equilibrium thermal speed $\vtpa$. The derivation of this closure relation will be presented in Appendix \ref{sec:appa} (see in particular Eq. (\ref{closmn}) to find the expressions for the various gyrofluid moments according to the quasi-static closure). We anticipate, however, two remarkable properties that this closure possesses. The first one is that the quasi-static closure relation turns out to be compatible with a Hamiltonian structure. The second one is that it allows for exact expressions, in terms of canonical Poisson brackets, for all the nonlinear terms in the gyrofluid equations, and in particular for those involving only gyroaveraged electromagnetic fields or potentials. This is not the case, to the best of our knowledge, for the previously derived reduced gyrofluid models.

The model assumes the presence of a strong magnetic guide field and evolves, for both electrons and ions, gyrocenter density fluctuations as well as velocity and temperature fluctuations referred to the direction parallel to the guide field. A dissipative variant of the model accounting for parallel Landau damping is then formulated through a Landau-fluid modeling of the parallel heat fluxes \citep{HP90,Ham92}. A four-field reduction assuming isothermality is presented as well. The resulting gyrofluid model retains ion and electron finite Larmor radius (FLR) corrections, electron inertia and parallel magnetic fluctuations,
enables  anisotropic  equilibrium temperatures and does not prescribe special restrictions on the ion or electron $\beta_{e,i}$ parameters, where $\beta_{e,i}$ indicates, for each species ($e$ for electrons and $i$ for ions), the ratio between the equilibrium kinetic pressure and the magnetic pressure exerted by the guide field. In this respect, this gyrofluid model differs from most of the presently available gyrofluid models, which require $\beta_{e,i} \ll 1$.

 In addition to the derivation of the model, we also carry out a detailed analysis of its linearized version. Predictions of the six-field model extended with Landau damping and of its four-field Hamiltonian reduction are compared at the linear level with those of the parent gyrokinetic model, by considering the dispersion relations of KAWs and slow waves (SWs), and also by analyzing the firehose  or the field-swelling \citep{Bas82,Bas84} instabilities. This latter instability, which requires the electron temperature
 transverse to the magnetic field larger than the longitudinal one, leads to a local increase in
 the transverse pressure, which  tends to make the magnetic field "swell" further locally, an effect which can  be important in producing magnetic reconnection. Its properties are rather subtle, especially when considering its effect on fast modes when the disturbance propagation is nearly perpendicular to the ambient magnetic field. We thus include an appendix summarizing the results on this instability that are relevant for the discussion of the present model.
As will be remarked in Sec. \ref{sec:complin}, for the non-dissipative case, the choice of the four-field model instead of the six-field model, for the comparison with the linear gyrokinetic theory, is due to the fact that a closure at an even order as in the four-field model, provides a better agreement with the Pad\'e approximant chosen for the electron response function in the quasi-static limit.

The investigation of the linear dispersion relations including temperature anisotropies is a first application of the model, mainly devoted to test its capability of reproducing results of the linear gyrokinetic theory in the appropriate regimes. We mention that the derivation of the model was motivated also by further physical applications of relevance to space plasmas and which will be part of subsequent works. These include the investigation of tearing instability in the presence of temperature anisotropy in a strong guide field regime, the influence of electron FLR effects and temperature anisotropy on inertial reconnection, or the effect of the coupling of Alfv\'en and compressible modes on the turbulence development.

The paper is organized as follows. In Sec. \ref{sec:parent} the gyrokinetic parent model is reviewed. In Sec. \ref{sec:gyrofluid} the six-field gyrofluid model is introduced and its Hamiltonian structure is presented. The two variants, corresponding to the Landau gyrofluid extension and to the Hamiltonian four-field reduction are also described.  Sec. \ref{sec:complin} is devoted to the comparison of the linearized versions of the two variants of the model with other linear theories.  We conclude in Sec. \ref{sec:concl} where we also mention the  interest of the present gyrofluid model for space plasma applications. At the end of the paper three Appendices are provided, presenting the derivation of the quasi-static closure relations from gyrokinetic theory, the derivation of the six-field gyrofluid model equations and a discussion of field-swelling instabilities, respectively.

\section{The  gyrokinetic parent model}  \label{sec:parent}

In order to derive a gyrofluid model only based on a quasi-static closure assumption, we consider as starting point the following set of gyrokinetics equations, which corresponds to the system provided by Eqs. (C58), (C60), (C66)-(C68) of \citet{Kun15} when collisions and equilibrium velocities are neglected and bi-Maxwellian distribution functions are chosen as equilibrium distribution functions for all the particle species. For simplicity, we specialize to the case of a plasma consisting of two species: electrons and one species of single ionized particles. The equations of the resulting gyrokinetic model  are given by:
\begin{align}
& \frac{\pa \ga}{\pa t}+\frac{c}{B_0}\left[ \mathcal{J}_{0s} \wphi - \frac{\vpar}{c} \mathcal{J}_{0s} \wapar +2 \frac{\mua B_0}{\qa}\mathcal{J}_{1s}\frac{\wbpar}{B_0} , \ga \right]
\nno \\
& +\vpar \frac{\pa}{\pa z}\left( \ga  +\frac{\qa}{\Tpa} \calfa\left( \mathcal{J}_{0s} \wphi - \frac{\vpar}{c} \mathcal{J}_{0s}\wapar +2 \frac{\mua B_0}{\qa} \mathcal{J}_{1s}\frac{\wbpar}{B_0}\right)\right)=0, \label{gyr}\\
&\sum_{s} \qa \int \dwa \, \mathcal{J}_{0s} \ga = \sum_{s} \frac{\qa^2 }{\Tpea} \int \dwa \,  \calfa \left( 1 - \mathcal{J}_{0s}^2 \right)  \wphi  \nno \\
&- \sum_s \qa \int \dwa \, 2 \frac{\mua B_0}{\Tpea} \calfa \mathcal{J}_{0s} \mathcal{J}_{1s} \frac{\wbpar}{B_0},  \label{qndim}\\
&\sum_s \qa \int \dwa \, \vpar \mathcal{J}_{0s}  \left( \ga -\frac{\qa}{\Tpa} \frac{\vpar}{c} \calfa \mathcal{J}_{0s}  \wapar\right) \nno \\
&= -\frac{c}{4 \pi} \lapp \wapar + \sum_s \frac{\qa^2}{m_s}\int \dwa \,\calfa \left( 1 - \frac{1}{\thea}\frac{\vpar^2}{\vtpa^2}\right)(1 - \mathcal{J}_{0s}^2) \frac{\wapar}{c},  \label{ampdim}\\
&\sum_s \frac{\bepea}{n_0} \int \dwa \, 2 \frac{\mua B_0}{\Tpea} \mathcal{J}_{1s} \ga= - \sum_s \frac{\bepea}{n_0} \frac{\qa}{\Tpea} \int \dwa \,  2 \frac{\mua B_0}{\Tpea} \calfa \mathcal{J}_{0s} \mathcal{J}_{1s}  \wphi  \nno \\
&-\left(2 + \sum_s \frac{\bepea}{n_0}\int \dwa \, \calfa \left( 2 \frac{\mua B_0}{\Tpea} \mathcal{J}_{1s} \mathcal{J}_{0s} \right)^2 \right) \frac{\wbpar}{B_0}. \label{amppedim}
\end{align}
The index $s \in \{e,i \}$ adopted above indicates the particle species, so that quantities labelled with $s$ refer to the electron species when $s=e$ and to the ion species when $s=i$.

In the system ({\ref{gyr}})-(\ref{amppedim}), the function $\ga$ is defined by
\beq  \label{defg}
\ga(x,y,z,\vpar,\mua,t)=\dfa(x,y,z,\vpar,\mua,t)+\frac{\qa}{\Tpa}\frac{\vpar}{c}\calfa (\vpar , \mua) \mathcal{J}_{0s} \wapar (x,y,z,t),
\eeq
where $\dfa$ is the perturbation of the gyrocenter distribution function for particles of species $s$, $\qa$ is the charge of these particles (so that $q_e=-e$ and $q_i=e$, with $e$ indicating the proton charge) and $m_s$ their mass. Furthermore, $c$ denotes the speed of light. The bi-Maxwellian equilibrium distribution function is given by 
\beq  \label{bimax}
\calfa(\vpar,\mua)= \left(\frac{m_{s}}{{2 \pi}}\right)^{3/2} \frac{n_0}{\Tpa^{1/2} \Tpea}\mathrm{e}^{-\frac{m_{s} \vpar^2}{2 \Tpa}-\frac{\mua B_0}{ \Tpea }},  
\eeq
where  $n_0$ is the uniform and constant equilibrium density, $\Tpa$ and $\Tpea$ are respectively the equilibrium temperatures of the $s$-th particle species parallel and in a plane perpendicular to an equilibrium magnetic guide field of amplitude $B_0$, directed along the $z$ direction of a Cartesian coordinate frame $\{x,y,z\}$. We suppose that the spatial domain of the system corresponds to a box $\cald=\{ (x,y,z): -L_x \leq x \leq L_x , -L_y \leq y \leq L_y, -L_z \leq z \leq L_z \}$, with $L_x$, $L_y$ and $L_z$ positive constants. All quantities of the system which depend on the spatial variables $x$, $y$ and $z$ are supposed to satisfy periodic boundary conditions on the domain $\cald$, so that they can be expanded in Fourier series. We indicated with $\vpar \in \mathbb{R}$ the velocity coordinate parallel to the guide field  and with $\mua=m_s v_\perp^2/(2 B_0) \in [0, +\infty)$  the magnetic moment of the particle of species $s$ in the unperturbed guide field, where $v_\perp$ corresponds to the velocity coordinate perpendicular to the guide field. We assume that all functions depending on $\vpar$ decay to zero as $\vpar \rightarrow \pm \infty$. Functions depending on  $\mua$  are assumed to tend to zero as $\mua \rightarrow +\infty$ and to be bounded at $\mua=0$. The coordinate $t\in [0, + \infty)$ refers to time. The expressions $\mathcal{J}_{0s}$ and $\mathcal{J}_{1s} $ are related to the standard gyroaverage operators for the species $s$ in Fourier space. Their definition can be introduced explicitly in the following way: adopting the  notation $\bx$ to indicate a point of coordinates $(x,y,z) \in \cald$ and, similarly, $\bk$ to indicate a point $(k_x, k_y, k_z)\in \mathscr{D}$, where $\mathscr{D}$ is the lattice defined by $\scrd=\{(2\pi l /(2 L_x), 2 \pi m/ (2 L_y) , 2 \pi n / (2 L_z)) : (l,m,n) \in \mathbb{Z}^3 \}$, we can consider a function $f : \cald \times [0, + \infty)\rightarrow \mathbb{R}$, periodic over $\cald$, so that it admits the Fourier representation $f(\bx ,t)=\sum _{\bk \in \mathscr{D}}  f_{\bk}(t) \exp(i \bk \cdot \bx)$.  The action of the operators $\mathcal{J}_{0s}$ and $\mathcal{J}_{1s}$ on the function $f$ is defined by
\begin{align}
&\mathcal{J}_{0s} f (\bx ,t)=\sum_{\bk \in \mathscr{D}} J_0 (\aal) f_{\bk}(t) \exp(i \bk \cdot \bx),\\
&\mathcal{J}_{1s} f (\bx ,t)=\sum_{\bk \in \mathscr{D}} \frac{J_1 (\aal)}{\aal} f_{\bk}(t) \exp(i \bk \cdot \bx),
\end{align}
where $J_0$ and $J_1$ indicate the zeroth and first order $J$ Bessel functions, respectively, $\aal=k_\perp \rsth$ is the perpendicular Larmor radius associated with the species $s$, with $k_\perp$ and $\rsth$ corresponding to the perpendicular wave number and  the Larmor radius of the particle of species $s$. The former is defined as $k_\perp=\sqrt{k_x^2+k_y^2}$ with $k_x=2 \pi l/(2 L_x)$, $k_y=2 \pi m/(2 L_y)$ for $l,m \in \mathbb{Z}$, while the latter is given by  $\rsth=v_\perp / \omega_{cs}$, where $\omega_{cs}=e B_0/ (m_s c)$ is the cyclotron frequency referred to the guide field and related to the particle of species $s$.

The leading order expression (up to second order terms in the perturbations) for the magnetic field is given by
\beq
\mathbf{B}(x,y,z,t)=\nabla \wapar (x,y,z,t) \times \hatz + (\wbpar ( x,y,z,t) + B_0)\hatz,
\eeq
where $\hatz$ is the unit vector along the $z$ direction, $\wapar$ (referred to as magnetic flux function) corresponds to the $z$ component of the magnetic vector potential and $\wbpar$ is the perturbation of the magnetic guide field, also referred to as parallel magnetic perturbation or parallel magnetic fluctuations. We remark that the guide field is assumed to be spatially homogeneous. This assumption is valid for a local description of space plasmas such as the solar wind, where the background magnetic field varies on scales so large that, in the local description, it can be assumed to be homogeneous. The situation would be different, for instance, in the case of tokamak plasmas. Indeed, gyrofluid models more oriented towards tokamak applications (as, for instance those of \cite{SH01,Mad13,Bri92,Sco10,WT_2012,Ker15}) take into account background magnetic inhomogeneities. The set of electromagnetic quantities involved in the system is completed by the electrostatic potential $\wphi=\wphi(x,y,z,t)$. In Eqs. (\ref{qndim})-(\ref{amppedim}) we adopted the symbol $\dwa=(2\pi B_0/m_s)d \mua d\vpar$ to indicate the volume element in space velocity. 

The parameters $\thea$ and $\bepea$ are defined by
\beq
 \thea=\frac{\Tpea}{\Tpa}, \qquad \bepea=8 \pi \frac{n_0 \Tpea}{B_0^2}
\eeq
 and measure, for each species $s$, the equilibrium temperature anisotropy and the ratio between equilibrium kinetic and magnetic pressure, respectively.

 Equation (\ref{gyr}) is the gyrokinetic equation related to the species $s$, whereas Eqs. (\ref{qndim})-(\ref{amppedim}) relate the gyrocenter distribution functions to electromagnetic quantities in the non-relativistic limit.
 In particular, Eq. (\ref{qndim}) corresponds to the quasi-neutrality relation, whereas Eqs. (\ref{ampdim}) and (\ref{amppedim}) descend form Amp\`ere's law projected along directions parallel and perpendicular to the guide field, respectively.

The gyrokinetic model (\ref{gyr})-(\ref{amppedim}) is valid for small perturbation of the equilibrium distribution function ( $\delta f$ approximation) and weak variations of the fields along the direction of the guide field, the equilibrium temperature anisotropy $\Theta_s$ and the parameter $\bepea$ of all the species, being kept finite in this asymptotics.  Further details about the derivation the regime of validity of the model and its derivation can be found in \citet{Kun15}. Its Hamiltonian structure, on the other hand, is presented in  \cite{Tas19}.
 
\section{The gyrofluid model}  \label{sec:gyrofluid}

We define the following gyrofluid moments:
\begin{align}
&\wns=\int \dwa \, \dfa , \qquad \wus=\frac{1}{n_0}\int \dwa \, \vpar \dfa,  \label{mom1}\\
&\wtps=\frac{\Tpa}{n_0} \int \dwa \, \left( \frac{\vpar^2}{\vtpa^2} -1\right) \dfa, \qquad \wtpes=\frac{\Tpea}{n_0} \int \dwa \, \left( \frac{\mua B_0}{\Tpea} -1 \right) \dfa,  \nno \\
&\qpars=\Tpa \vtpa \int \dwa \, \left(\frac{\vpar^3}{\vtpa^3} - 3 \frac{\vpar}{\vtpa}\right) \dfa. \label{mom3}
\end{align}
For each particle species, the fields $\wns$ and $\wus$ represent the fluctuations of the gyrocenter densities and parallel fluid velocities, respectively. On the other hand, $\wtps$ and $\wtpes$ correspond to the fluctuations of the gyrofluid temperatures defined with respect to the parallel and perpendicular gyrocenter velocities, respectively, whereas $\qpars$ indicates the gyrocenter parallel heat flux fluctuations. In defining the parallel temperature and heat flux fluctuations, we introduced the constant $\vtpa=\sqrt{\Tpa / m_s}$, corresponding to the parallel thermal velocity associated with the species $s$.

We intend to derive a gyrofluid model by taking moments of the gyrokinetic equation (\ref{gyr}) and by imposing a closure relation derived from the quasi-static linear theory. In particular, the gyrofluid model, accounting for equilibrium temperature anisotropy, should be able, in the limit of vanishing finite Larmor radius effects, to reproduce the field-swelling instability criterion of \citet{Bas84}. We restrict to the evolution of the first three moments referring to the parallel direction. Therefore, the resulting gyrofluid model should evolve the following six fields: $\widetilde{N}_e, \widetilde{N}_i, \widetilde{U}_e, \widetilde{U}_i, \widetilde{T}_{\parallel e}$ and $\widetilde{T}_{\parallel i}$.  
Also, we show that the model conserves the total energy and, moreover, that it possesses a noncanonical Hamiltonian structure, as is  the case for the parent gyrokinetic model \citep{Tas19}. We refer to such model as to the 6-field gyrofluid (GF6) model. 

For the sake of the comparison, carried out in Sec. \ref{sec:complin}, of the linear gyrofluid theory with the linear gyrokinetic theory we also consider an extension of GF6 accounting for a Landau-fluid closure, analogously to that discussed in  \citet{HP90,Ham92,Snyder97,PS07,PS15,Tas18}. This variant of the model, which we denote as GF6L, differs from GF6 for the expression of the parallel heat flux fluctuations $\qpars$. Therefore, in order to avoid some redundancy in the exposition, in the following we present the model equations leaving $\qpars$ unspecified and we will subsequently indicate the corresponding expressions for $\qpars$ leading to GF6 and GF6L, respectively. The closure leading to GF6L, in particular, will be given in Sec. \ref{ssec:variants}.

A further variant of the model, denoted as GF4, will also be considered in Sec. \ref{sec:complin}. This model evolves only the four fields $\widetilde{N}_e, \widetilde{N}_i, \widetilde{U}_e$ and $\widetilde{U}_i$ and represents a minimal Hamiltonian model, derived from the quasi-static closure, capable to reproduce the field-swelling instability criterion. This variant of the model will be introduced in Sec. \ref{ssec:variants}.

The six-field system, both in the Hamiltonian (GF6) and dissipative (GF6L) versions, can be written, in a dimensionless form, as
\begin{align}
&\frac{\pa N_s}{\pa t}+[\gamus\phi + \sgn(\qa) \taupa 2 \gamds \bpar, N_s] - [\gamus \apar, U_s ]   +\frac{\pa U_s}{\pa z}=0,  \label{conta}  \\
&\frac{\pa }{\pa t}\left( \frac{m_s}{m_i} U_s  + \sgn(\qa) \gamus\apar\right)\nno \\
&+\left[ \gamus \phi +\sgn(\qa) \taupa 2 \gamds \bpar, \frac{m_s}{m_i} U_s + \sgn(\qa) \gamus\apar \right] \label{ua}  \\
&-\frac{\taupa}{\thea}[\gamus \apar , N_s +\Tps] \nno \\
& +\frac{\pa}{\pa z} \left( \sgn(\qa) \gamus \phi + \taupa 2 \gamds\bpar+\frac{\taupa}{\thea}(N_s +\Tps) \right)=0, \nno \\
&\frac{\pa \Tps}{\pa t}+[\gamus\phi + \sgn(\qa) \taupa 2 \gamds \bpar, \Tps]  \nno \\
&- [\gamus \apar, 2 U_s + \qps]   +\frac{\pa}{\pa z}( 2 U_s + \qps)=0,  \label{tempa}\\
&\sum_s \left(\sgn( \qa ) \gamus N_s + (1 -\thea) \gamzs \frac{\phi}{\taupa}+(\thea \gamus^2  -1)\frac{\phi}{\taupa}  \right. \nno \\
& \left. + \sgn (\qa ) (1-\thea)(\gamzs-\gammus) \bpar +\sgq \thea 2 \gamus \gamds \bpar\right)=0, \label{qna}\\
&\lapp \apar= \frac{\bepe}{2}\sum_s \left(\frac{m_i}{m_s}\left(1-\frac{1}{\thea}\right)(1-\gamzs) \apar -\sgn (\qa) \gamus U_s \right),  \label{ampa}\\
&\bpar=-\frac{1}{2}\sum_s \bepa \left(2 \gamds N_s+(1-\thea)(\gamzs-\gammus)\sgq\frac{\phi}{\taupa}  \right. \nno \\
&\left. + \thea 2 \gamus\gamds\sgq	\frac{\phi}{\taupa}+2  (1-\thea)(\gamzs - \gammus)\bpar + \thea 4 \gamds^2  \bpar \right).  \label{ampea}
\end{align}
In Eqs. (\ref{conta})-(\ref{ampea}), we adopted the following normalized quantities
\begin{align}
& N_s=\frac{\wns}{n_0}, \qquad U_s=\frac{\wus}{\cspe},  \label{norm1}\\
& \Tps=\frac{\wtps}{\Tpa}, \qquad \qps=\frac{\qpars}{n_0 \Tpa \cspe}, \qquad  \label{norm2}\\
&\phi= \frac{e \wphi}{\Tpee}, \qquad \apar=\frac{\wapar}{B_0 \rs}, \qquad \bpar=\frac{\wbpar}{B_0},  \label{norm3}\\
&x=\frac{\bar{x}}{\rs}, \qquad y=\frac{\bar{y}}{\rs}, \qquad z=\frac{\bar{z}}{\rs}, \qquad  t=\omega_{ci}{\bar{t}},  \label{norm4}
\end{align}
where the quantities with overbars in Eq. (\ref{norm4}) are the dimensional spatial and time coordinates.
In Eqs. (\ref{norm1})-(\ref{norm4}) the quantities\footnote{Note that, according to a customary notation, in the symbols $\cspe$ and $\rs$, the subscript $s$ is to indicate sonic quantities and not the particle or gyrocenter species.}
\beq
\cspe=\sqrt\frac{\Tpee}{m_i}, \qquad \rs=\frac{\cspe}{\omega_{ci}}
\eeq
were introduced, which indicate the sound speed and the sonic Larmor radius, respectively, based on the perpendicular equilibrium temperature.

The parameter
\beq
\taupa=\frac{\Tpea}{\Tpee},
\eeq
for $s=e,i$, on the other hand, measures the ratio of the equilibrium perpendicular temperatures. 

In Eqs. (\ref{conta})-(\ref{ampea}) we introduced the operator  $b_s= -\lapp \rthpes^2$, with $\lapp$ denoting the Laplacian relatively to the transverse variables and
\beq
\rthpes=\frac{1}{\omega_{cs}}\sqrt{\frac{\Tpea}{m_s}}
\eeq
indicating the perpendicular thermal Larmor radius associated with the species $s$. 

The canonical bracket $[ \, , \, ]$, on the other hand, is defined as $[f,g]=\pa_x f  \pa_y g - \pa_y f \pa_x g$, for two functions $f$ and $g$. 

The gyroaverage operators $\gamus$, $\gamds$, $\gamzs$, $\gammus$ in Eqs. (\ref{conta})-(\ref{ampea}) can in turn be expressed in terms of the operators  
\begin{align}
&\gamu(\ba)=\mathrm{e}^{-\ba/2}, \qquad \gamd(\ba)=\frac{\mathrm{e}^{-\ba/2}}{2},  \label{op1}\\
&\Gamma_0(\ba)=I_0( \ba)\mathrm{e}^{-\ba}, \qquad \Gamma_1(\ba)=I_1(\ba)\mathrm{e}^{-\ba}, \label{op2}
\end{align}
which have to be intended as Fourier multipliers whose symbols are obtained by replacing $b_s$ by $k_\perp^2 \rho^2_{th \perp s}$. We make this statement more precise, as an example, in the case of the operator $\gamui$, referred to the ion species.  The expression $\gamui f $  is defined by $\gamui f (\bx ,t)=\sum_{\bk \in \mathscr{D}} \gamu (b_i) f_{\bk}(t) \exp(i \bk \cdot \bx)=\sum_{\bk \in \mathscr{D}} \exp(-k_\perp^2 \rthpei^2/2) f_{\bk}(t) \exp(i \bk \cdot \bx)$, for a function $f$ periodic in space. Analogous expressions are valid for the other gyroaverage operators. In Eq. (\ref{op2}), in particular, the symbols $I_0$ and $I_1$ indicate the modified Bessel functions $I$ of order zero and one, respectively. 

The expressions for the operators $\gamus$ and $\gamds$ given in Eq. (\ref{op1}) correspond to those present in \citet{Bri92} and follow from assuming that the perturbation of the distribution function can be written as
\beq  \label{expf}
\dfa (x,y,z,\vpar,\mua,t)=\calfa (\vpar ,\mua)\sum_{m,n =0}^{+\infty}\frac{1}{\sqrt{m !}} H_m\left(\frac{\vpar}{\vtpa}\right)L_n\left(\frac{\mua B_0}{\Tpea}\right)f_{{mn}_s} (x,y,z,t),
\eeq
where $H_m$ and $L_n$ indicate the Hermite and Laguerre polynomials, respectively, of order $m$ and $n$, with $m$ and $n$ non-negative integers. The functions $f_{{mn}_s}$ are coefficients of the expansion and are related to the moments of $\dfa$, with respect to Hermite polynomials in $\vpar / \vtpa$ and Laguerre polynomials in $\mua B_0 / \Tpea$. Indeed, from the orthogonality properties of Hermite and Laguerre polynomials, the following relation holds:
\beq
f_{{mn}_s}=\frac{1}{n_0 \sqrt{ m !}}\int \dwa \,  H_m  \left(\frac{\vpar}{\vtpa}\right)L_n\left(\frac{\mua B_0}{\Tpea}\right) \dfa.
\eeq
It can be useful, in particular,  to write explicitly the following relations between the lowest order normalized moments and the quantities $f_{{mn}_s}$:
\begin{align}
& f_{{00}_s}=N_s, \qquad f_{{10}_s}=\sqrt{\frac{\thea}{\taupa}}\sqrt{\frac{m_s}{m_i}}U_s,   \label{momf0} \\
& f_{{20}_s}=\frac{1}{\sqrt{2}}T_{\|s}, \qquad f_{{01}_s}=-T_{\perp s},  \label{momf}\\
& f_{{30}_s}=\frac{1}{\sqrt{6}}\sqrt{\frac{\thea}{\taupa}}\sqrt{\frac{m_s}{m_i}}\qps,  \label{momf3}
\end{align}
where, in Eq. (\ref{momf}), we introduced the normalized gyrocenter perpendicular temperature fluctuations $T_{\perp s}=\wtpes/\Tpea$.

As above anticipated, in the system (\ref{conta})-(\ref{ampea}) we temporarily left unspecified the expression for the parallel heat flux fluctuations $Q_{\|s}$. In order to obtain the model GF6, the infinite hierarchy of gyrofluid equations following from the parent gyrokinetic model (\ref{gyr})-(\ref{amppedim}) is closed by imposing relations obtained by computing the gyrocenter moments other than $N_s$, $U_s$ and $T_{\|s}$  from a
 linearization of the parent  gyrokinetic system about a homogeneous equilibrium, in the quasi-static limit. In the case of $\qps$, this leads  (consider Eqs. (\ref{momf3}) and (\ref{closmn})) to
 \beq  \label{closgf6}
 \qps=0, \qquad s=e,i.
 \eeq
 The model GF6 is thus obtained by inserting the relation (\ref{closgf6}) into the system (\ref{conta})-(\ref{ampea}).
  Details on the derivation of the closure relations originated from the quasi-static assumption can be found in Appendix \ref{sec:appa}. 
  A remarkable property of this closure is that it  leads to the annihilation of all the contribution of the higher order moments in the gyrofluid equations. 
 
 We find it useful to provide also a reformulation of the evolution equations (\ref{conta})-(\ref{tempa}), which should help putting in evidence the physical nature of the  terms contributing to the evolution of the various fields. Eqs. (\ref{conta})-(\ref{tempa}) can indeed be rewritten as
 \begin{align}
 &\frac{\pa N_s}{\pa t}+\us\cdot\nabla N_s+ \gpars U_s=0, \label{contshort}\\
 &\frac{m_s}{m_i}\frac{\pa U_s}{\pa t}+\frac{m_s}{m_i}\us\cdot\nabla U_s-\sgn (q_s)\epars + \gpars \mathcal{P}_s=0,  \label{momshort}\\
 &\frac{\pa \Tps}{\pa t}+\us\cdot\nabla \Tps+ 2\gpars U_s=-\gpars \qps, \label{tempshort}
\end{align}
where the parallel gradient operator $\gpars$ is defined, for each species $s$, by
\beq
\gpars f=-[\gamus \apar, f ] + \frac{\pa f}{\pa z},
\eeq
for a function $f$.
From the formulation (\ref{contshort})-(\ref{tempshort}) it emerges that the gyrocenter density, parallel momentum and parallel temperature fluctuations are all advected, in the perpendicular plane, by the incompressible velocity field
\beq  \label{adv}
\us=\hz\times\nabla\left(\gamus \phi + \sgn (q_s) \taupa 2 \gamds \bpar\right).
\eeq
Such velocity field  includes a first contribution, associated with $\gamus \phi$, which corresponds to the usual $\mathbf{E}\times\mathbf{B}$ drift (based on the gyroaveraged electrostatic potential), ubiquitous in low-$\beta$ gyrofluid models such as those discussed in \cite{WT_2012}, \cite{SH01}, \cite{Ker15} and \cite{Wae09}. When higher $\beta$ values are allowed, however, the perpendicular advection acquires a further contribution due to the parallel magnetic perturbations, as it transpires from Eq. (\ref{adv}). We remark that, similarly to the $\mathbf{E}\times\mathbf{B}$ contribution, also the latter contribution does not vanish in the limit $b_s \rightarrow 0$ of negligible FLR corrections.    This is a consequence of the fact that, in the presence of parallel magnetic perturbations, gyrocenter density fluctuations differ from particle density fluctuations, even in the absence of FLR corrections \citep{Bri92}.

From the continuity equation (\ref{contshort}), we see that the evolution of gyrocenter density fluctuations has also a source in the last term on the left-hand side, which is due to the compressibility of the gyrocenter mean velocity along the direction of the magnetic field.

From Eq. (\ref{momshort}) we see that parallel momentum, in addition to be advected by $\us$, evolves due to the term $-\sgn (q_s)\epars$, where $\epars$ is defined by
\beq
\epars=-\frac{\pa \gamus \apar}{\pa t}-[ \gamus \phi , \gamus \apar]-\frac{\pa \gamus \phi}{\pa z}.
\eeq
Such term represents the force exerted by the gyroaveraged electric field, along the direction of the magnetic field. A further source for the parallel momentum is due to the term $\gpars \mathcal{P}_s$, where
\beq
\mathcal{P}_s=\taupa \left(2 \gamds \bpar +\frac{1}{\thea}(N_s+\Tps)\right).
\eeq
This terms is associated with the parallel component of the divergence of an anisotropic pressure tensor.

The parallel temperature equation (\ref{tempshort}) has, on its left-hand side, the same structure of the continuity equation (\ref{contshort}). We just remark the presence of the coefficient $2$ multiplying $\gpars U_s$. This coefficient, of course, follows directly from taking the second order moment, in Hermite polynomials for the normalized parallel velocity, of the gyrokinetic equation (\ref{gyr}), as discussed in Appendix \ref{sec:appb}. However, as pointed out by \cite{Ker15}, in the presence of background magnetic curvature such coefficient, in general, has to be adjusted in order to obtain a Hamiltonian structure. Finally, we consider the term on the right-hand side of Eq. (\ref{tempshort}), associated with the parallel heat flux. If the expression for $\qps$ is chosen according to the quasi-static closure, i.e. imposing the relation (\ref{closgf6}), this term vanishes and the system, as will be shown in Sec. \ref{ssec:hamgf6}, is Hamiltonian. On the other hand, if the Landau fluid closure (\ref{landau}) is chosen, this term acts as a sink and the system is not energy-conserving.

 \subsection{Hamiltonian structure of GF6}  \label{ssec:hamgf6}

 The quasi-static closure (\ref{closgf6}) leading to GF6, allows the resulting model to be cast in Hamiltonian form. In particular, it can  be verified by direct computation that, when the electromagnetic fluctuations $\phi$, $\apar$ and $\bpar$ can be expressed in terms of the dynamical variables $N_e$, $N_i$, $M_e$ and $M_i$ (where we introduced the short-hand notation $M_s=(m_s/m_i)U_s+\sgn(\qa)\gamus\apar$ to indicate the parallel canonical momenta), by making use of  the relations (\ref{qna})-(\ref{ampea}), the evolution equations (\ref{conta})-(\ref{tempa}) complemented by Eq. (\ref{closgf6}), can be written in the Hamiltonian form 
 \beq  \label{hamform}
 \frac{\pa N_s}{\pa t}=\{N_s , H\}, \qquad \frac{\pa M_s}{\pa t}=\{M_s , H\}, \qquad \frac{\pa \Tps}{\pa t}=\{\Tps , H\}, \qquad  s=e,i.
 \eeq
 In Eq. (\ref{hamform}) $H$ is the Hamiltonian functional
 \begin{align}
 &H(N_e,N_i,M_e,M_i, \Tpe, \Tpi)= \nno \\
 &\frac{1}{2}\sum_{s} \int d^3 x \, \left( \frac{\taupa}{\thea} N_s^2 + \frac{m_i}{m_s} M_s^2+(\sgn (\qa) \gamus \phi + 2 \taupa \gamds \bpar )N_s  \right. \nno \\
 &\left. -\sgn(\qa) \frac{m_i}{m_s} \gamus \apar M_s + \frac{\taupa}{\thea} \frac{\Tps^2}{2}\right), \label{ham}
 \end{align}
 and $\{ \, , \, \}$ is a noncanonical Poisson bracket given by
  \begin{align}
  &\{F,G\}=- \sum_s \int d^3 x \, \left[  \sgn (\qa) \left(N_s \left( [F_{N_s} , G_{N_s}]+\frac{\taupa}{\thea}\frac{m_s}{m_i}[F_{M_s} , G_{M_s}] + 2 [F_{\Tps}, G_{\Tps}]\right) \right. \right. \nno \\
  & \left. \left. +M_s([F_{M_s} , G_{N_s}] +[F_{N_s} , G_{M_s}] +2 ([F_{M_s} , G_{\Tps}]+[F_{\Tps} , G_{M_s}]))\right.\right. \nno\\
  &\left. \left. + \Tps \left( \frac{\taupa}{\thea}\frac{m_s}{m_i} [F_{M_s} , G_{M_s}] + [F_{N_s} , G_{\Tps}] + [F_{\Tps} , G_{N_s}] + [F_{\Tps} , G_{\Tps}] \right)\right)\right. \nno\\
  & \left. +F_{N_s} \frac{\pa G_{M_s}}{\pa z} + F_{M_s}\frac{\pa G_{N_s}}{\pa z} + 2F_{\Tps} \frac{\pa G_{M_s}}{\pa z} + 2 F_{M_s}\frac{\pa G_{\Tps}}{\pa z}  \right],  \label{pb}
  \end{align}
  for two functionals $F$ and $G$. For details about the noncanonical Hamiltonian formulation of fluid models one can refer, for instance, to \citet{Mor98}. In Eq. (\ref{pb}) the subscripts on the functionals indicate functional derivatives. In order to verify the formulation (\ref{hamform}) it is convenient to remark that, from Eq. (\ref{ham}), one obtains
  \beq  \label{funcder}
  H_{N_s}=\frac{\taupa}{\thea}N_s+\sgn (\qa) \gamus \phi + \taupa 2 \gamds \bpar, \qquad H_{M_s} = U_s, \qquad H_{\Tps}=\frac{\taupa}{\thea}\frac{\Tps}{2}.
  \eeq
  In order to derive the relations (\ref{funcder}) we made use of the formal symmetry of the operators $\gamus$ and $\gamds$, i.e. $\int d^3x \, f \gamus g = \int d^3 x \, g \gamus f$ and $\int d^3x \, f \gamds g = \int d^3 x \, g \gamds f$ for two functions $f$ and $g$, as well as of the formal symmetry of the linear operators in terms of which one can express $\phi, \bpar$ and $\apar$ in terms of $N_s$ and $M_s$ through Eqs. (\ref{qna})-(\ref{ampea}). We did not provide the explicit expression for such operators which can, however, be obtained considering the representation in Fourier series of the fields involved, following the procedure discussed in \cite{Tas19}.
  
  The Hamiltonian functional (\ref{ham}) is a conserved quantity for the dynamics and corresponds to the total energy.
  
  In Eq. (\ref{pb}), the sum of all the terms with $s=e$ is a Poisson bracket in its own right. Similarly, all the terms with $s=i$ form a Poisson bracket.  The sum of these two contributions is a direct sum of Poisson brackets which is in turn a Poisson bracket verifying in particular the Jacobi identity. The Poisson brackets referring to the electron and ion quantities correspond to those already discussed in other Hamiltonian reduced fluid models. The reader can in particular refer to \citet{Tas15} and \citet{Ker15} for the verification of the Jacobi identity for brackets of such form and for a discussion of the corresponding Casimir invariants. The model GF6, in the two-dimensional limit when the dependence on the $z$ coordinate is suppressed, can also be cast in the form of a system of advection of equations for Lagrangian invariants, as is the case for several other reduced fluid and gyrofluid models (see, e.g. \cite{Ker15,Wae09,WT_2012,Gra15,Tas15,Tas19,Sch94}).

\medskip
\noindent
{\bf Remark:}
The gyroaverage operators $\gamu$ with a form different from (\ref{op1}) have been proposed in the literature (see, e.g.  \citet{Dor93,SH01,Sco10}), when the expansion (\ref{expf}) is not assumed, and are frequently adopted.  In particular,  $\gamu (b_s)=\Gamma_0^{1/2} (b_s)$ was shown to provide better agreement with the linear theory at large $b_s$ \citep{Dor93}. We point out, however, that many important features of the  model (\ref{conta})-(\ref{ampea}),  such as the total energy conservation and the Hamiltonian structure, are guaranteed whatever the form of the operators $\gamu$ and $\gamd$ of Eq. (\ref{op1}) is, provided these operators are linear and formally symmetric, in the sense defined above. This is in particular the case  with  $\gamu (b_s)=\Gamma_0^{1/2} (b_s)$. These issues are also discussed by \citet{Man18}.

\subsection{Variants of the model}\label{ssec:variants}

\subsubsection{Six-field model with Landau closure (GF6L)}  
   
 The variant GF6L of the six-field gyrofluid model, accounting for Landau damping, corresponds to the system (\ref{conta})-(\ref{ampea}) with $\qps$ given by
 \beq  \label{landau}
\qps =-2 \alpha_s \mathcal{L} T_{\parallel s}, \qquad s=e,i.
\eeq
 In Eq. (\ref{landau}) we introduced the constant  $\alpha_s=(2/\pi)^{1/2}(m_i/m_s)^{1/2}$. The operator $\cal{L}$ holds for the Landau damping operator. Its modeling in the nonlinear regime is discussed in \citet{Tas18}. In the linear approximation,   it reduces to the negative of the Hilbert transform in
the direction of the ambient magnetic field (here taken in the $z$ direction). The presence of this Landau operator in reduced fluid models breaks the Hamiltonian structure by violating energy and Casimir conservation (\cite{Tas18,Gra20}). Its purpose, on the other hand, is to introduce terms that allow the linear dispersion relation of the gyrofluid model to reproduce that of the parent gyrokinetic model.

\subsubsection{Four-field model with quasi-static closure (GF4)}  \label{sssec:gf4}
The second variant GF4 is obtained by retaining the evolution equations for $N_s$ and $M_s$ and imposing the quasi-static closure on the parallel temperature fluctuations $\Tps$. Considering Eqs. (\ref{momf}) and (\ref{closmn}), this amounts to setting 
\beq  \label{closgf4}
\Tps=0, \qquad s=e,i.
\eeq
The resulting model reads
\begin{align}
&\frac{\pa N_s}{\pa t}+[\gamus\phi + \sgn(\qa) \taupa 2 \gamds \bpar, N_s] - [\gamus \apar, U_s ]   +\frac{\pa U_s}{\pa z}=0,  \label{contgf4}  \\
&\frac{\pa }{\pa t}\left( \frac{m_s}{m_i} U_s  + \sgn(\qa) \gamus\apar\right)\nno \\
&+\left[ \gamus \phi +\sgn(\qa) \taupa 2 \gamds \bpar, \frac{m_s}{m_i} U_s + \sgn(\qa) \gamus\apar \right] \label{ugf4}  \\
&-\frac{\taupa}{\thea}[\gamus \apar , N_s ]  +\frac{\pa}{\pa z} \left( \sgn(\qa) \gamus \phi + \taupa 2 \gamds\bpar+\frac{\taupa}{\thea}N_s  \right)=0, \nno \\
&\sum_s \left(\sgn( \qa ) \gamus N_s + (1 -\thea) \gamzs \frac{\phi}{\taupa}+(\thea \gamus^2  -1)\frac{\phi}{\taupa}  \right. \nno \\
& \left. + \sgn (\qa ) (1-\thea)(\gamzs-\gammus) \bpar +\sgq \thea 2 \gamus \gamds \bpar\right)=0, \label{qnf4}\\
&\lapp \apar= \frac{\bepe}{2}\sum_s \left(\frac{m_i}{m_s}\left(1-\frac{1}{\thea}\right)(1-\gamzs) \apar -\sgn (\qa) \gamus U_s \right),  \label{ampf4}\\
&\bpar=-\frac{1}{2}\sum_s \bepa \left(2 \gamds N_s+(1-\thea)(\gamzs-\gammus)\sgq\frac{\phi}{\taupa}  \right. \nno \\
&\left. + \thea 2 \gamus\gamds\sgq	\frac{\phi}{\taupa}+2  (1-\thea)(\gamzs - \gammus)\bpar + \thea 4 \gamds^2  \bpar \right) \label{ampef4}
\end{align}
and corresponds to taking Eqs. (\ref{conta}), (\ref{ua}), (\ref{qna}), (\ref{ampa}) and (\ref{ampea}) of GF6 with $\Tps=0$. Its Hamiltonian structure is given by the Hamiltonian
 \begin{align}
 &H(N_e,N_i,M_e,M_i)= \nno \\
 &\frac{1}{2}\sum_{s} \int d^3 x \, \left( \frac{\taupa}{\thea} N_s^2 + \frac{m_i}{m_s} M_s^2+(\sgn (\qa) \gamus \phi + 2 \taupa \gamds \bpar )N_s  \right. \nno \\
 &\left. -\sgn(\qa) \frac{m_i}{m_s} \gamus \apar M_s \right) \label{hamf4}
 \end{align}
 and by the Poisson bracket
  \begin{align}
  &\{F,G\}=- \sum_s \int d^3 x \, \left[  \sgn (\qa) \left(N_s \left( [F_{N_s} , G_{N_s}]+\frac{\taupa}{\thea}\frac{m_s}{m_i}[F_{M_s} , G_{M_s}] \right) \right. \right. \nno \\
  & \left. \left. +M_s([F_{M_s} , G_{N_s}] +[F_{N_s} , G_{M_s}] ) \right)+F_{N_s} \frac{\pa G_{M_s}}{\pa z} + F_{M_s}\frac{\pa G_{N_s}}{\pa z} \right].
  \end{align}
 If one neglects electron FLR effects (i.e. $b_e \rightarrow 0$), parallel magnetic perturbations (i.e. $\bpar =0$), equilibrium temperature anisotropies (i.e. $\Theta_e=\Theta_i=1$) and
 sets $\gamu (b_i)=\Gamma_{0}^{1/2} (b_i)$ (i.e. takes the alternative form of the ion gyroaverage operator mentioned in the above remark in Sec. \ref{ssec:hamgf6}), GF4 reduces to the Hamiltonian gyrofluid model of \cite{WT_2012}, the latter taken in the limit of vanishing magnetic curvature and equilibrium density gradients. We note, however, that, once that the quasi-static closure relations are determined, as in Eq. (\ref{closmn}), all the terms in GF4 (and likewise for GF6), are determined exactly. In particular no approximations of the gyroaverage operators (unlike, for instance, in \cite{WT_2012} and \cite{Sco10}) are carried out. Terms involving gyroaverage operators are determined exactly also in \cite{Bri92}, but without making use of the quasi-static closure. As a result, in \cite{Bri92}, nonlinear terms involving more than one gyroaverage operator do not result in having the single canonical bracket structure  (as is the case in GF4, for instance,  with the term $\sgn (\qa) [ \gamus \phi ,\gamus \apar]$ appearing in the second line of Eq. (\ref{ugf4})) and    which is crucial for determining the Lie-Poisson Hamiltonian structure. This property, which follows from the quasi-static closure, differentiates the models presented in our paper also from its closest predecessors, i.e. the gyrofluid models constructed with the technique recently presented in \cite{Tas19}. The latter gyrofluid models, in fact, are also Hamiltonian and account for equilibrium temperature anisotropy, but adopt a different closure. Namely, all gyrofluid moments involving finite powers of the magnetic moment $\mua$ (e.g. the perpendicular temperature fluctuations) are set equal to zero. This allows for a Hamiltonian structure but the terms involving gyroaverage operators are not all determined exactly from taking the moments  of the gyrokinetic equations. For this reason, we think that, for situations where the quasi-static assumption $|\omega/(k_z \vtpa)| \ll1$ is satisfied, the models presented in this paper might be preferable to the models described in \cite{Tas19}. 
   
 \section{Comparison with the linearized gyrokinetic parent model}  \label{sec:complin}
 \subsection{KAWs dispersion relation}
 
 We first discuss the KAW dispersion relation as predicted by the gyrofluid models, with and without Landau damping
 (GF6L and GF4 respectively), in comparison with the predictions of the linearized parent model \citep{Kunz18} which identifies with the low-frequency kinetic theory described in \citet{KPS12}. This latter dispersion relation involves the plasma response function $R(\zeta_s)$ of the particles of species $s$, which is related to the corresponding plasma dispersion function $Z(\zeta_s)$ by $R(\zeta_s) = 1 + \zeta_s Z(\zeta_s)$ with $\zeta_s = \omega/(2^{1/2}|k_z|v_{th_{\| s}})$.
 Different  Pad\'e approximants $R_{ij}(\zeta_e)$ (for which we follow the notations of \citet{Hunana19})  are used to estimate the electron response function. 
 
In order to test the validity of the quasi-static assumption that affects the form of the FLR terms, independently from the effects resulting from Landau damping, we are  led to compare the prediction of the 4-field gyrofluid model GF4 with the gyrokinetic dispersion relation obtained by choosing for the electron plasma response function, the function $R_{21}(\zeta_e)=1/(1-2\zeta_e^2)$ (model denoted GKNL). This choice directly results from the assumption ${T_{{\|e}}}=0$ (see Eq. (B7) of \citet{PS07}) and ignores  electron Landau damping\footnote{Note that, choosing a closure that sets equal to zero a higher even order moment  in the hierarchy of parallel moments would only improve the matching with kinetic theory at very large $\zeta_e$ but not in the zero-$\zeta_e$ limit. Differently, closures at an odd order
are not consistent with the quasi-static assumption, as they rather correspond to an adiabatic regime.}. Predictions of the GF6L model will, on the other hand, be compared with the full gyrokinetic dispersion relation, referred to as GK. In the latter description, we use for the electron Pad\'e approximant the function $R_{86}(\zeta_e)$, which shows an excellent agreement with the exact response function. For the ions,  we use in all the cases the  Pad\'e approximant $R_{20}(\zeta_i)=1/(1-2\zeta_i^2-i\pi^{1/2}\zeta_i)$ (or $R_{21}(\zeta_i)$ in the absence of Landau damping). Higher order Pad\'e approximants give almost identical results. 
\begin{table} 
\begin{center}
\caption{Acronyms and corresponding models.}
\begin{tabular}{cc}
\hline
{Acronym} & {Model}   \\
\hline
 GF6 & Hamiltonian six-field gyrofluid model with quasi-static closure \\
  & (Eqs. (\ref{conta})-(\ref{ampea}) with $\qps$ given by Eq. (\ref{closgf6})) \\
 \hline
 GF6L &  Six-field gyrofluid model with Landau closure \\
 
  & (Eqs. (\ref{conta})-(\ref{ampea}) with $\qps$ given by Eq. (\ref{landau}))  \\
  \hline
  GF4 & Hamiltonian four-field gyrofluid model with quasi-static closure \\
   & (Eqs. (\ref{contgf4})-(\ref{ampef4}))\\ 
   \hline
   GK & Full gyrokinetic dispersion relation (from \citet{Kunz18})\\
   \hline
   GKNL & Gyrokinetic dispersion relation with no Landau damping\\
   \hline
   BC84 & Asymptotic model by \cite{Bas84} (Eq. (\ref{disp4}))\\
 \hline
   DK & Full drift-kinetic dispersion relation (Eq. (\ref{disp}))\\
   \hline
   DKNL & Drift-kinetic  dispersion relation with no  Landau damping\\
\end{tabular}

\label{tab1}
\end{center}
\end{table}

\begin{figure}
 	\centerline{
 		\includegraphics[width=0.31\textwidth]{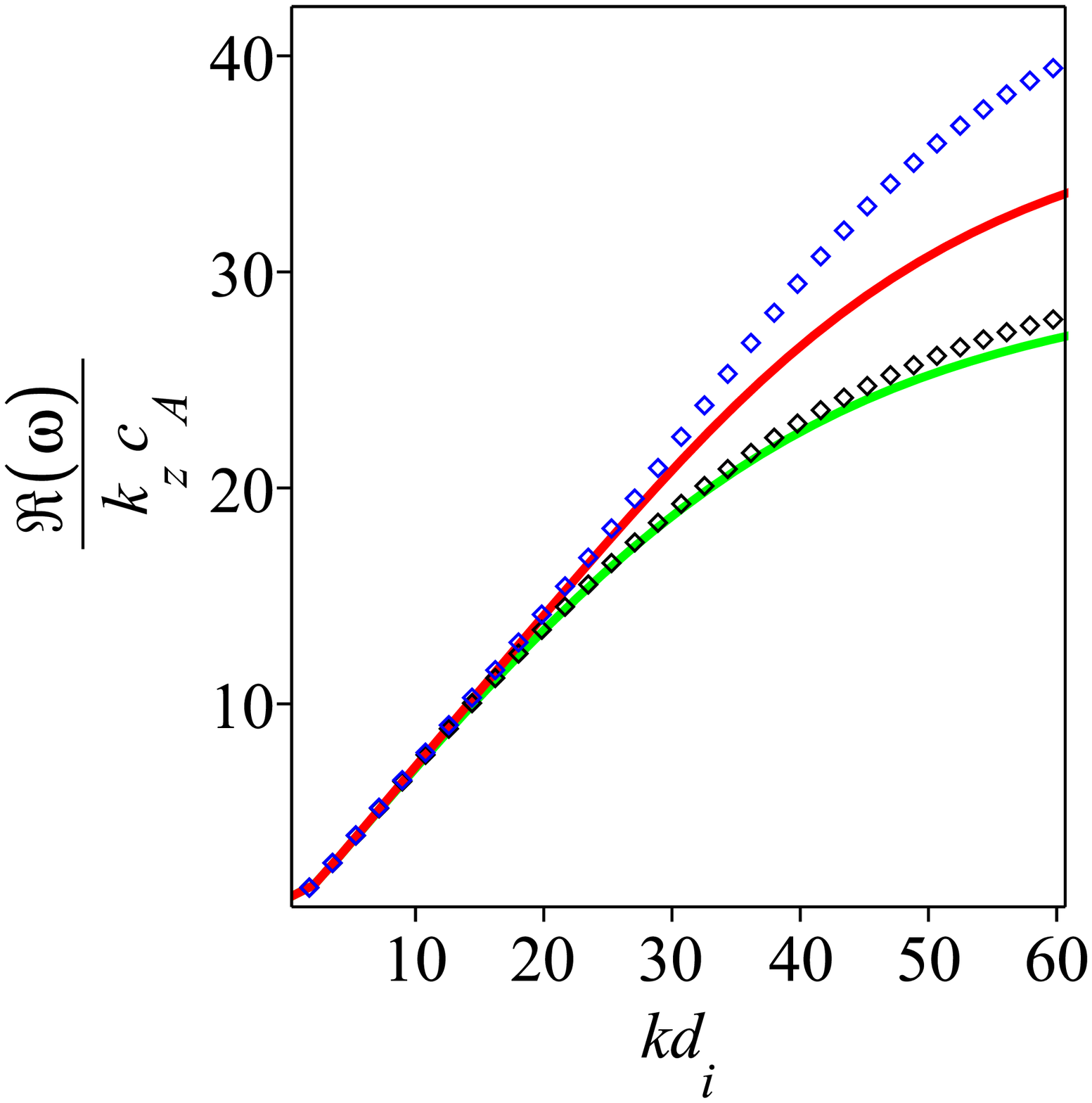}
 		\includegraphics[width=0.31\textwidth]{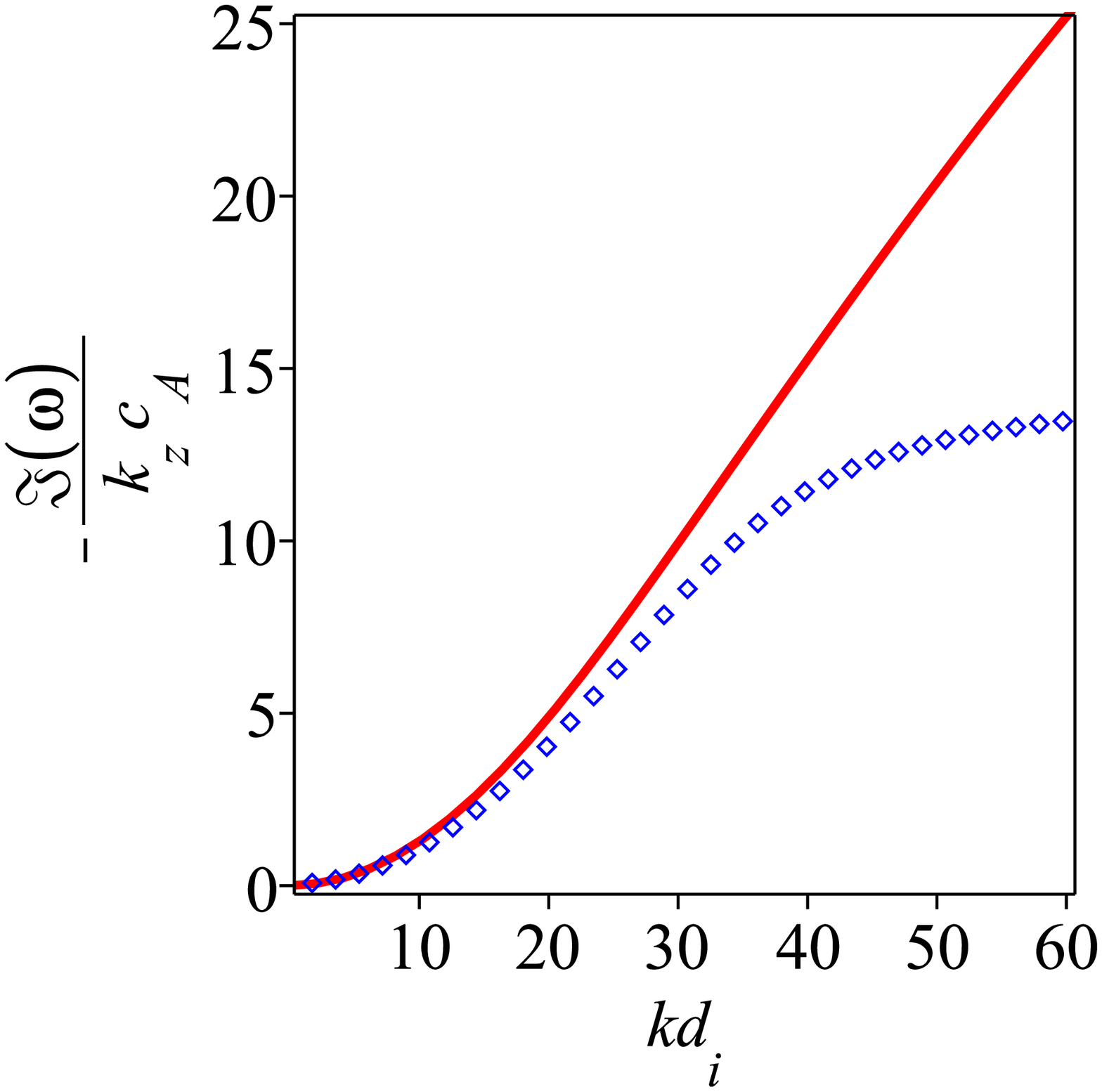}
 		\includegraphics[width=0.31\textwidth]{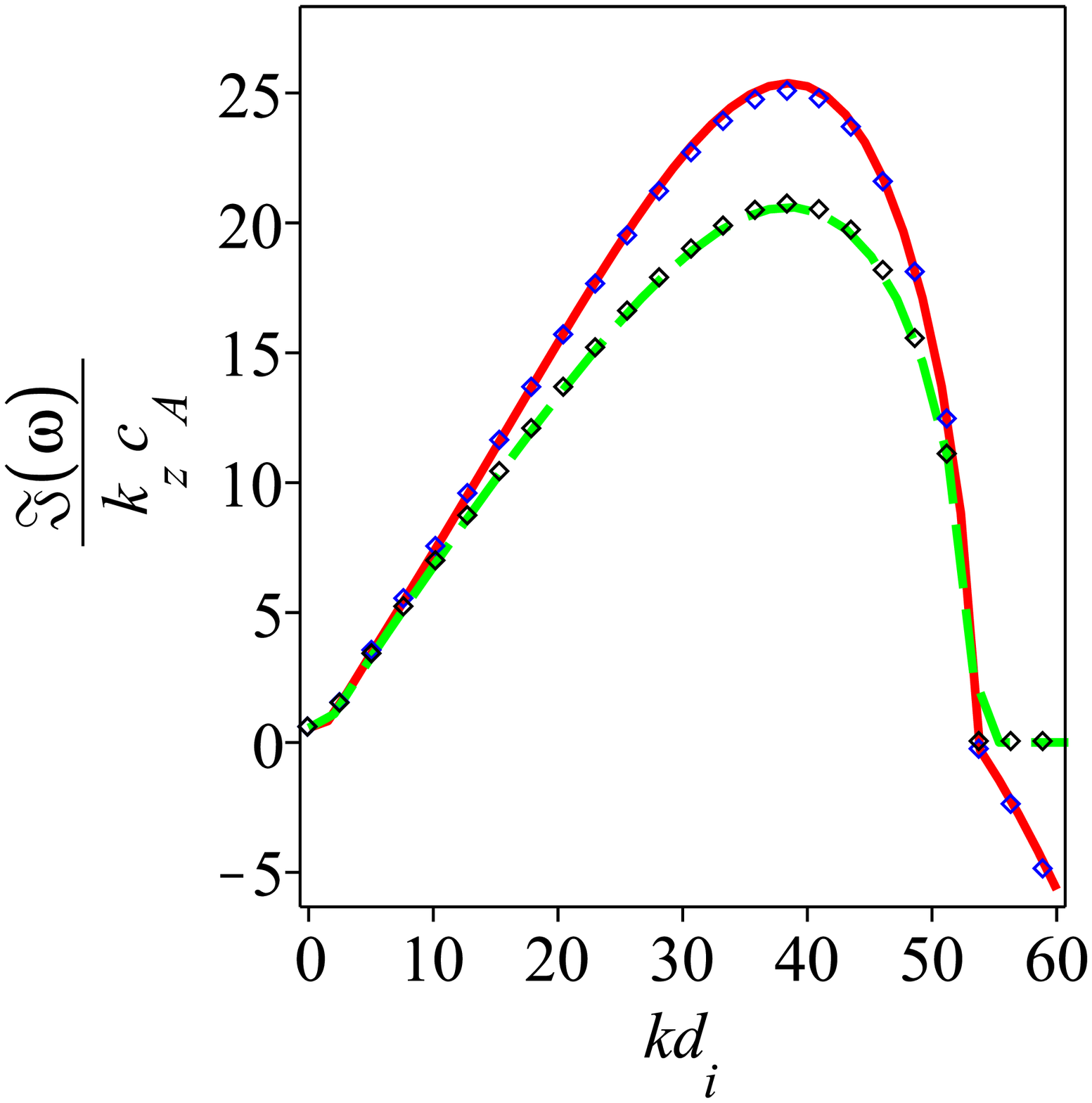}}
 \caption{Left: Normalized real part of the KAW frequency $\omega/(k_z c_A)$ for $\beta_{\perp e}=1$, $\tau_\|=1$, $\Theta_e=1$, $\alpha=89^\circ$, as a function of $k d_i$, calculated using GF4 (black diamond symbols) and GF6L (blue diamond symbols) models (respectively without and with Landau damping)  and with the gyrokinetic dispersion relations GKNL (green solid line) and GK (red solid line). Middle:  Damping rate from the GK theory  (red solid line) and from  the GF6L model (blue diamond symbols). Right: Normalized growth rate of the firehose instability as a function of $k d_i$, for  $\beta_{\perp e}=0.5$, $\tau_\|=10^{-3}$,   $\Theta_e=0.16$, $\alpha=89^\circ$,  as predicted  by  GF4 and GF6L models, together with GKNL and GK dispersion relations, using the same graphic conventions.}
 \label{fig:kaw}
 \end{figure}

Because we are making use of several acronyms to refer to the different adopted models, we summarized them  in Table \ref{tab1}. 
 In all the examples presented in this Section, $\Theta_i=1$.
 Introducing the ion to electron parallel temperature ratio at equilibrium $\tau_\| = {T_{0_{\|i}}}/{T_{0_{\|e}}}=\tau_{\perp_ i}\Theta_e/{\Theta_i}$
and denoting by $\alpha$ the angle between the wave vector and the ambient magnetic field (propagation angle), we compare, in
 Fig. \ref{fig:kaw} (left),  the (real) normalized KAW frequency $\omega/(k_z c_A)$ for $\beta_{\perp e}=1$, $\tau_\|=1$, $\Theta_e=1$,  $\alpha=89^\circ$ as a function of $k d_i$
 	(where $k=\sqrt{k_\perp^2+k_z^2}$ and $d_i$ is the ion inertial length defined by $d_i = c_A/\omega_{ci}$, with $c_A =\sqrt{B_0/(4\pi m_i n_0)}$ the Alfv\'en velocity),  calculated using GF4 (black diamond symbols), with the GKNL prediction (green solid line). 
 An excellent agreement is found, the slight deviation appearing as $kd_i>40$ probably resulting from the failure of the quasi-static approximation used in the calculation of all the moments starting from the temperature. 
 For the same plasma parameters, the case with Landau damping is displayed as a red solid line for GK and blue diamond symbols for GF6L, with a good agreement up to scales $kd_i \lessapprox 20$.
 The damping rate is displayed with the same symbols in Fig. \ref{fig:kaw} (middle). Interestingly, the prediction of GF6 (not shown) departs from GKNL significantly at all the scales. Indeed, in GF6, the parallel temperature fluctuations that obey a dynamical equations with zero heat flux do not approach a quasi-static dynamics. Such a non-dissipative odd-order closure rather fits an adiabatic regime (see footnote in this Section). Landau damping is requested to ensure convergence to a quasi-static regime. In that sense, GF4 is preferable to GF6 for addressing a non-dissipative problem, the interest of the latter model being mainly to provide a framework where Landau damping can be supplemented to an otherwise Hamiltonian description.
 
 \subsection{Firehose instability}
 Figure  \ref{fig:kaw} (right) displays for $\beta_{\perp e}=0.5$, $\tau_\|=10^{-3}$, $\alpha=89^\circ$, $\Theta_e=0.16$, the growth rate of the firehose instability as a function of $kd_i$ predicted by  GF4 and GF6L (black and blue diamond symbols respectively), together with the GKNL and GK dispersion relations (green and red solid lines). In all the cases, the agreement is excellent, the quenching of the instability being in particular well reproduced.
 
 The agreement found between the KAWs dispersion relations predicted by the gyrofluid models and the gyrokinetic theory, even when limited to scales such that $kd_i \lessapprox 20$, and despite a  large value of  $\zeta_i = \sqrt{2\Theta_e/(\beta_{\perp_e}\tau_\|)}\;\omega/(|k_z|c_A)$,
 can be attributed to the fact that, at least within the linear theory, ion acoustic and kinetic Alfv\'en waves remain essentially decoupled. The influence of the ion closure relation on the KAW properties thus remains limited. Deviations from the gyrokinetic theory at small scales are mostly due to the fact that when reaching these scales  
 $\zeta_e=\sqrt{m_e/m_i}\sqrt{2\Theta_e/\beta_{\perp_e}}\;\omega/(|k_z|c_A)$
 becomes non-negligible.

\subsection{SW dispersion relation and field swelling instability}
 
\begin{figure}
 	\centerline{
 		\includegraphics[width=0.48\textwidth]{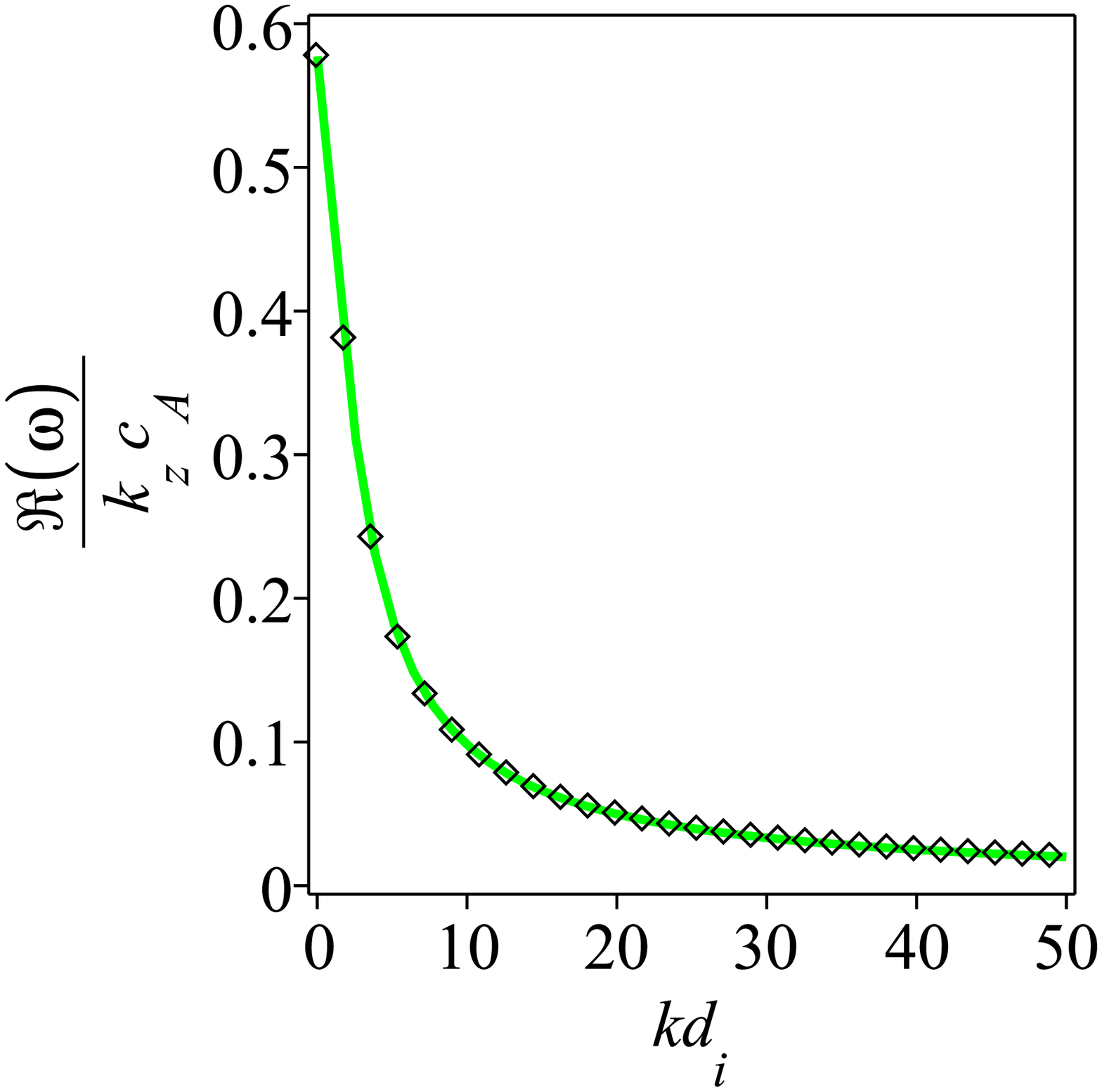}
 		\includegraphics[width=0.48\textwidth]{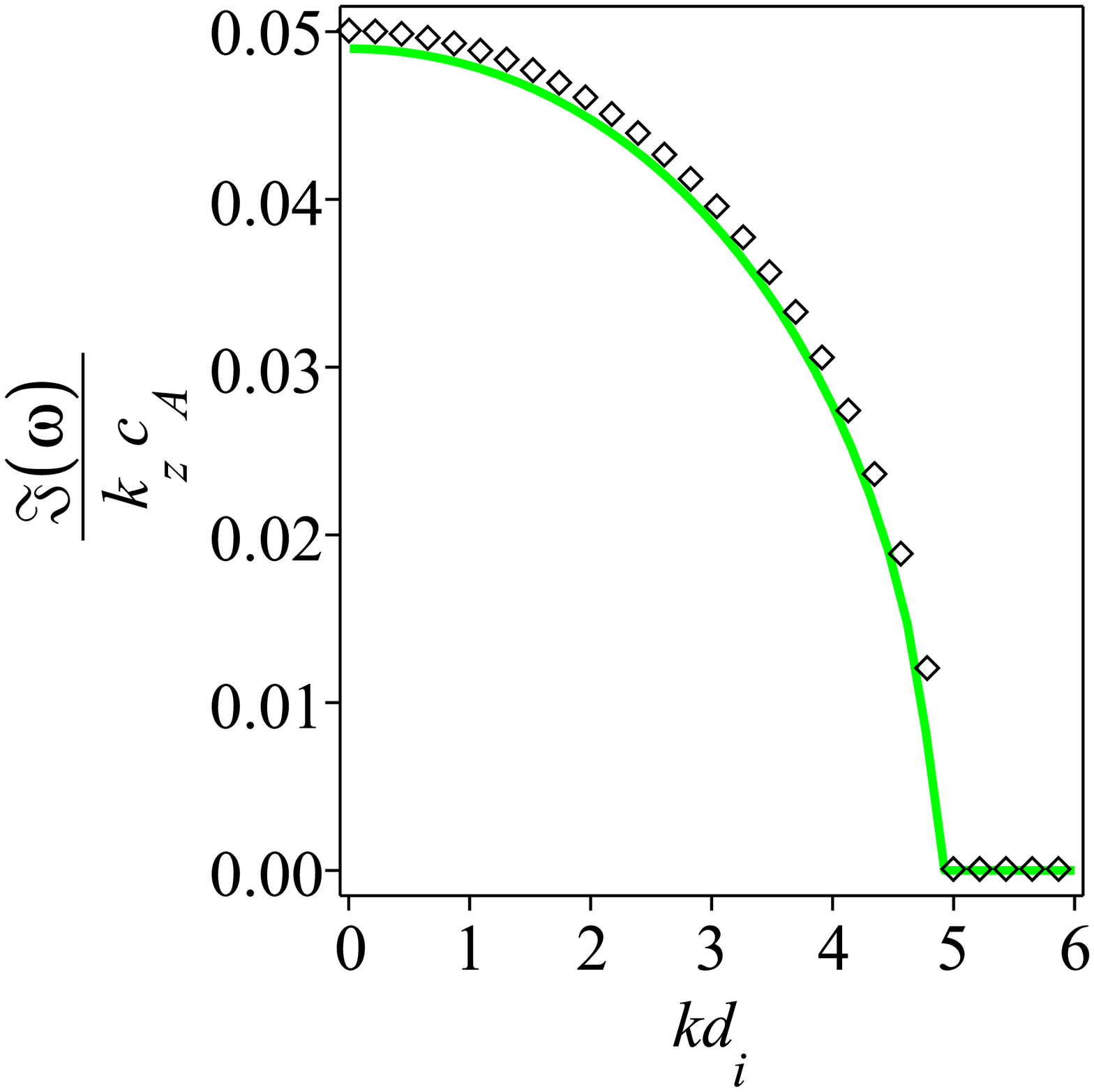}}
 	\caption{Left: Normalized slow-wave frequency versus $kd_i$ in regimes where the ions are cold ($\tau_\|=10^{-5}$), with $\beta_{\perp e}=1$ and $\alpha=89^\circ$, as predicted by the GF4 model and the GKNL dispersion relation. Right: normalized growth rate of the swelling instability for $\tau_\|=10^{-5}$ and $\Theta_e=2.01$. Same conventions as in Fig. \ref{fig:kaw} are used.}
 	\label{fig:sw2-versus-k}
 \end{figure}

 Figure \ref{fig:sw-versus-k} concerns a similar comparison in the case of the field swelling instability discussed in Appendix \ref{App:swelling}, for $\beta_{\perp e}=1$, $\tau_\|=1$, $\alpha=89^\circ$, $\Theta_e=2.2$ (left). For both the cases with and without Landau damping, the relatively good agreement found at large scale between the gyrofluid theories and the gyrokinetic ones deteriorates at smaller scales. The stabilization scale is however correctly captured. This discrepancy is associated with a value of $\zeta_i$, which is not small enough (in this case, $\zeta_e$ remains reasonably small). We show in Fig. \ref{fig:sw-versus-k} (right) with  $\Theta_e=2.41$, again with $\beta_{\perp e}=1$ and  $\alpha=89^\circ$, that a much better agreement can be found  when ions are hotter ($\tau_\|=50$),  which prescribes a small $\zeta_i$. The case with cold ions ($\tau_\|=10^{-5}$), for which the ion dynamics is decoupled, is displayed in Fig. \ref{fig:sw2-versus-k}, showing an even better agreement. The left panel displays the real part of the slow wave for $\Theta_e=1$, obtained with the GKNL model  (solid green line) or the GF4 model (diamond symbols). The right panel shows the growth rate of the field-swelling instability for $\Theta_e=2.01$ (keeping unchanged the other parameters).

 \begin{figure}
 	\centerline{
 	\includegraphics[width=0.48\textwidth]{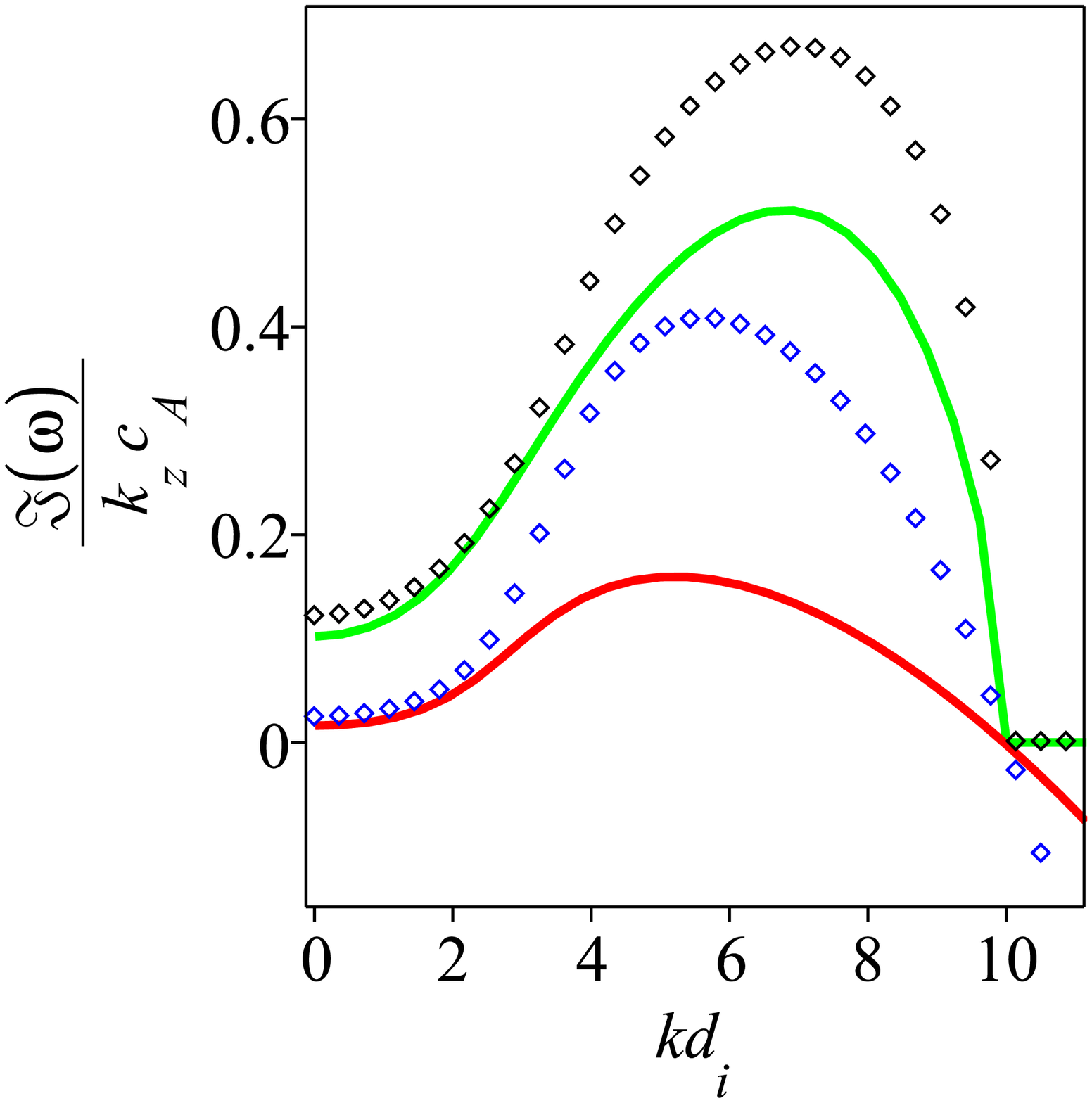}
 	\includegraphics[width=0.48\textwidth]{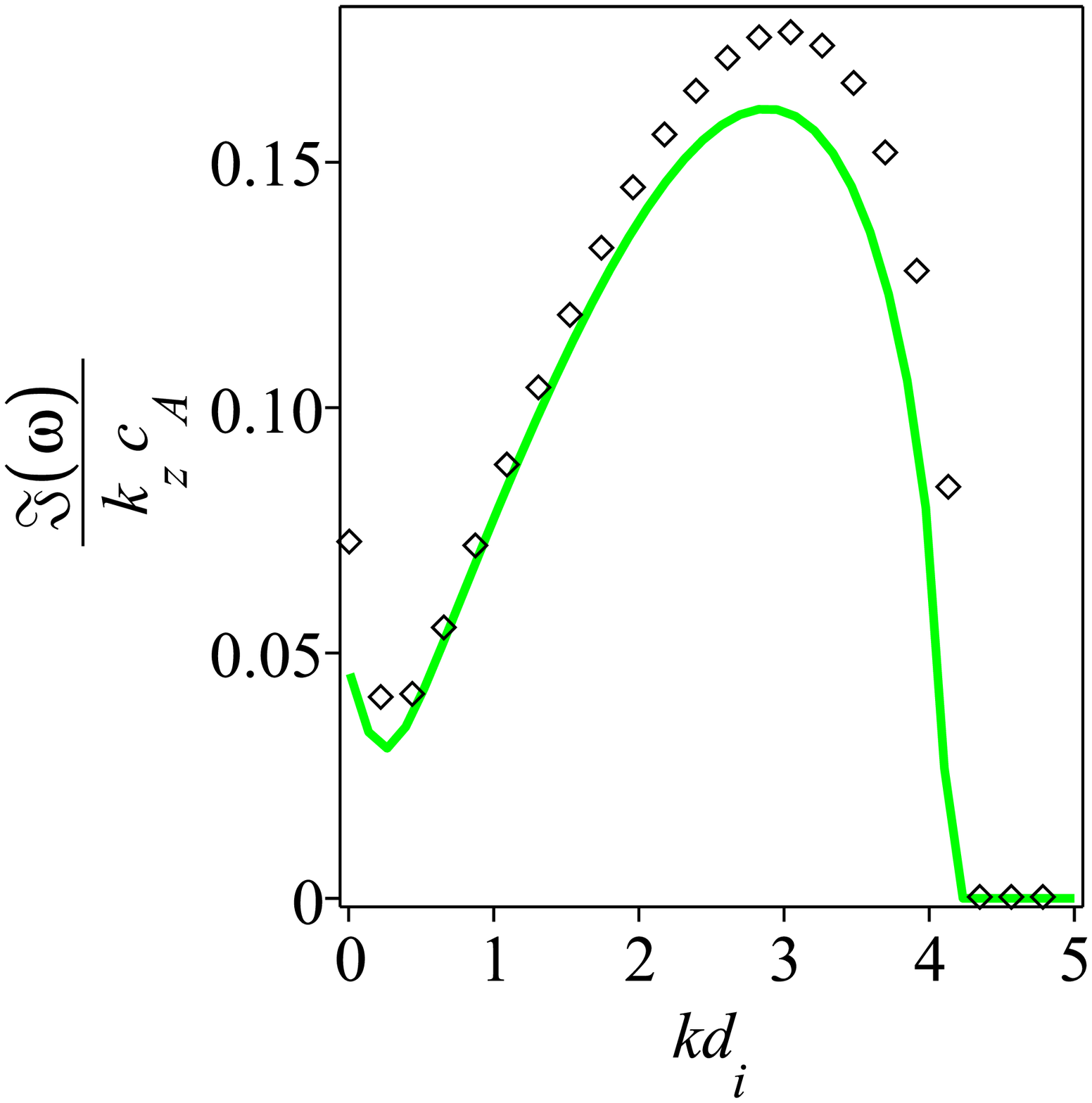}}
\caption{Normalized growth rate of the field swelling instability versus $kd_i$ for $\alpha=89^\circ$ and  $\beta_{\perp e}=1$,
Left:   Predictions of GF4 and GF6L, compared with those of  GKNL and GK respectively,  for $\tau_\|=1$ and  $\Theta_e=2.2$.
Right: Prediction of GF4 compared with that of GKNL,  for $\tau_\|=50$ and $\Theta_e=2.41$ . Same graphic conventions as in Fig. \ref{fig:kaw} are used.}

 	\label{fig:sw-versus-k}
 \end{figure}
 
 \section{Conclusion} \label{sec:concl}
 We derived a 6-field Hamiltonian gyrofluid model, referred to as GF6, retaining the gyrocenter density, the  parallel velocity and temperature fluctuations for each species, under the sole assumption that all the other gyrocenter moments are calculated from the quasi-static linear kinetic theory. Such an assumption on the closure turns out to yield exact expressions for all the terms of the model, without, in particular, requiring approximated expressions for the terms involving gyroaverage operators. Nonlinear terms involving more than one gyroaverage operator, in particular, appear in the form of a single canonical bracket, which naturally lets the model fit in the class of Hamiltonian models with a Lie-Poisson structure. The model accounts for equilibrium temperature anisotropy and also retains both ion and electron FLR corrections, electron inertia and parallel magnetic fluctuations. In a  variant of the model (GF6L) parallel Landau damping is retained through a Landau-fluid modelization of the gyrocenter parallel heat fluxes. A second variant of the model (GF4) is obtained by prescribing parallel isothermality, which still falls in the frame of the quasi-static closure and allows for a Hamiltonian formulation.
The  comparison of the dispersion relations of  KAWs and SWs predicted from GF6L or GF4, with those derived from the parent gyrokinetic theory where the plasma response function is replaced by a Pad\'e approximant, provides an estimate  of  the maximal  transverse wavenumber beyond which the phase velocity of the corresponding wave is too large compared with the electron (in the former case) or ion (in the latter case) parallel thermal velocity for consistency with a closure condition based on a quasi-static assumption. It turns out that  the agreement extends to transverse scales significantly smaller than the ion Larmor radius in the case of KAWs, mostly because, at least at the linear level, SWs and KAWs are essentially decoupled, making the influence of the ion closure relation on the KAW properties relatively limited. This situation contrasts with the case of the SWs for which the dispersion relation is accurately reproduced only at scales larger than a significant fraction of the ion Larmor radius. Under these conditions, the model reproduces the instabilities induced by temperature anisotropy, such as firehose or field-swelling instabilities. It should  nevertheless  be noted  that, as it assumes small perturbations of an equilibrium state,  the model does not permit evolution of the mean temperatures, an effect usually considered as contributing  efficiently to the saturation of these instabilities. 
 The subcritical nonlinear regime is however expected to be accurately described. The model will in particular be most useful for studying the coupling of  KAWs with  SWs which can
 generate large-scale parametric decay instabilities at small $\beta_e$, a regime especially relevant in the regions of the solar wind relatively close to the Sun explored by space missions such
 as Parker Solar Probe or Solar Orbiter.
 
  In general, to the best of our knowledge, our gyrofluid model is the only one, at the present moment, possessing the following features, which could make it a valuable tool for local investigations of basic plasma phenomena of interest for space plasmas:  it accounts for equilibrium temperature anisotropies as well as parallel magnetic perturbations; it reproduces, in a rather wide range of values of parameters, compatible with the quasi-static closure, quantitative features of known kinetic linear dispersion relations; the model equations, and in particular the terms involving FLR corrections, are calculated exactly, unlike other gyrofluid models which adopt truncations or approximations of such terms;  it possesses a Hamiltonian structure.
  \newpage
  
 \begin{appendix}
 
 \section{Derivation of closure relations from the gyrokinetic linear theory} \label{sec:appa}
 
 We consider the linearization of the gyrokinetic system (\ref{gyr})-(\ref{amppedim}) about the  equilibrium state $\ga=0$ (or, equivalently, $\dfa=0$), with $\wphi=\wapar=\wbpar=0$.
 The resulting linear system can be written in the form
\begin{align}
& \frac{\pa }{\pa t}\left(\dfa +\frac{\qa}{\Tpa}\frac{\vpar}{c}\calfa  \mathcal{J}_{0s}\wapar \right)  +\vpar \frac{\pa}{\pa z}\left( \dfa  +\frac{\qa}{\Tpa} \calfa\left( \mathcal{J}_{0s}  \wphi  +2 \frac{\mua B_0}{\qa} \mathcal{J}_{1s}\frac{\wbpar}{B_0}\right)\right)=0, \label{lingyr}\\
&\sum_{s} \qa \int \dwa \, \mathcal{J}_{0s}  \dfa = \sum_{s} \frac{\qa^2 }{\Tpea} \int \dwa \,  \calfa \left( 1 - \mathcal{J}_{0s}^2 \right)  \wphi  \nno \\
&- \sum_s \qa \int \dwa \, 2 \frac{\mua B_0}{\Tpea} \calfa \mathcal{J}_{0s}  \mathcal{J}_{1s} \frac{\wbpar}{B_0},  \label{linqndim}\\
&\sum_s \qa \int \dwa \, \vpar \mathcal{J}_{0s}  \dfa \nno \\
&= -\frac{c}{4 \pi} \lapp \wapar + \sum_s \frac{\qa^2}{m_s}\int \dwa \,\calfa \left( 1 - \frac{1}{\thea}\frac{\vpar^2}{\vtpa^2}\right)(1 - \mathcal{J}_{0s}^2) \frac{\wapar}{c},  \label{linampdim}\\
&\sum_s \frac{\bepea}{n_0} \int \dwa \, 2 \frac{\mua B_0}{\Tpea} \mathcal{J}_{1s} \dfa= - \sum_s \frac{\bepea}{n_0} \frac{\qa}{\Tpea} \int \dwa \,  2 \frac{\mua B_0}{\Tpea} \calfa \mathcal{J}_{0s}  \mathcal{J}_{1s}  \wphi  \nno \\
&-\left(2 + \sum_s \frac{\bepea}{n_0}\int \dwa \, \calfa \left( 2 \frac{\mua B_0}{\Tpea} \mathcal{J}_{1s}\right)^2 \right) \frac{\wbpar}{B_0}, \label{linamppedim}
\end{align}

	where we adopted the same notation with the tilde symbol, that we used in Eqs. (\ref{gyr})-(\ref{amppedim}), to indicate the dynamical variables $\dfa$ of the linearized system and the field perturbations $\wphi, \wapar$ and $\wbpar$.

We introduce the following Fourier series representation:
\begin{align}
&\dfa(\bx , \vpar , \mua ,t)=\sum_{\bk \in \scrd} \dfk (\vpar ,\mua)\mathrm{e}^{i (\bk\cdot \bx -\omega t)}, \qquad \wphi(\bx ,t)=\sum_{\bk \in \scrd} \phik \mathrm{e}^{i (\bk\cdot \bx -\omega t)},  \label{fourier1}\\
&  \wapar(\bx ,t)=\sum_{\bk \in \scrd} \ak \mathrm{e}^{i (\bk\cdot \bx -\omega t)}, \qquad \wbpar(\bx ,t)=\sum_{\bk \in \scrd} \bpark \mathrm{e}^{i (\bk\cdot \bx -\omega t)},  \label{fourier2}
\end{align}
 with $\omega \in \mathbb{C}$ indicating the complex frequency.
 
 For any given $\bk \in \scrd$, from Eq. (\ref{lingyr}) we obtain the relation
 \begin{align}
& \dfk=\frac{\zs}{(\vpar /\vtpa)- \zs}\frac{\vpar}{c}\frac{\qa}{\Tpa}\calfa J_0(\aal) \ak  \nno \\
&- \frac{1}{(\vpar /\vtpa)-\zs}\frac{\vpar}{\vtpa} \calfa \left( \frac{\qa}{\Tpa} J_0 (\aal) \phik+ \frac{2 \mua B_0}{\Tpa} \frac{J_1 (\aal)}{\aal} \frac{\bpark}{B_0}\right),  \label{lin1}
 \end{align}
 where $\zs=\omega/(k_z \vtpa)$.
 
 We consider now the quasi-static limit $\vert \zs \vert \ll 1$. In this limit, the relation (\ref{lin1}) reduces to
 \beq  \label{qslin}
 \dfk = - \calfa \left( \frac{\qa}{\Tpa} J_0 (\aal) \phik+ \frac{2 \mua B_0}{\Tpa} \frac{J_1 (\aal)}{\aal} \frac{\bpark}{B_0}\right).
 \eeq
 In the following, we make use of the relation (\ref{qslin}), derived from the linear theory, in order to determine the closure relations to insert in the hierarchy of nonlinear gyrofluid equations. For this purpose,
 we adopt the Hermite-Laguerre expansion of Eq. (\ref{expf}) for the perturbation of the distribution function in the linearized system and we write
 \beq  \label{expfk}
  \dfa(\bx , \vpar , \mua ,t)=\calfa (\vpar ,\mua)\sum_{m,n =0}^{+\infty}\frac{1}{\sqrt{m !}} H_m\left(\frac{\vpar}{\vtpa}\right)L_n\left(\frac{\mua B_0}{\Tpea}\right)f_{{mn}_s} (\bx,t)
 \eeq
 with
 \beq  \label{fmnFourier}
f_{{mn}_s} (\bx ,t)= \sum_{\bk \in \scrd} \fks \mathrm{e}^{i (\bk\cdot \bx -\omega t)}
\eeq
in Fourier representation. From Eqs. (\ref{expfk}), (\ref{fmnFourier}), using the orthogonality relations for Hermite and Laguerre polynomials, one obtains
\beq   \label{fmnk}
\fks=\frac{1}{n_0 \sqrt{ m !}}\int \dwa \, \dfk H_m  \left(\frac{\vpar}{\vtpa}\right)L_n\left(\frac{\mua B_0}{\Tpea}\right).
\eeq
Inserting the relation (\ref{qslin}) into Eq. (\ref{fmnk}) and using the orthogonality relations for Hermite polynomials, one has
\beq  \label{closk}
\fks=-\delta_{m0}\left( G_{1n} \frac{\qa}{\Tpa} \phik + 2 \thea G_{2n} \frac{\bpark}{B_0}\right),
\eeq
where the operators $G_{1n}$ and $G_{2n}$ are defined by
\begin{align}
&G_{1n}=\frac{2 \pi B_0}{m_s} \int d \mua \, \mathsf{f}_{eq_s}(\mua) L_n \left( \frac{\mu_s B_0}{\Tpea}\right) \jo ,   \label{gyroop1}\\
& G_{2n}=\frac{2 \pi B_0}{m_s} \int d \mua \, \mathsf{f}_{eq_s}(\mua) L_n \left( \frac{\mu_s B_0}{\Tpea}\right) \frac{\mua B_0}{\Tpea} \frac{\ju}{\aal},   \label{gyroop2}
\end{align}
 with
 \beq 
 \mathsf{f}_{eq_s}(\mua)=\frac{m_s}{2 \pi \Tpea}\mathrm{e}^{-\frac{\mua B_0}{\Tpea}}.
 \eeq
 Explicit expressions for the operators $G_{1n}$ and $G_{2n}$ can be found by computing the integrals in Eqs. (\ref{gyroop1}) and (\ref{gyroop2}), which yields
 \begin{align}
 & G_{1n} (b_s) = \frac{\mathrm{e}^{-b_s /2}}{n!}\left(\frac{b_s}{2}\right)^n, \qquad n\geq 0,   \label{G1s}\\
 & G_{20} (b_s)= \frac{\mathrm{e}^{-b_s /2}}{2}, \quad G_{2n} (b_s) =-\frac{\mathrm{e}^{-b_s/2}}{2} \left( \left(\frac{b_s}{2}\right)^{n-1} \frac{1}{(n-1)!}-\left(\frac{b_s}{2}\right)^{n} \frac{1}{n !}\right), \qquad n \geq 1.  \label{G2s}
 \end{align}
 In order to obtain the expressions (\ref{G1s})-(\ref{G2s}) use was made of the orthogonality of Laguerre polynomials as well as of the relations \citep{Sze75}
 \begin{align}
 & \jo=\mathrm{e}^{-b_s/2}\sum_{n=0}^{+\infty} \frac{L_n\left( \frac{\mu_s B_0}{\Tpea}\right)}{n !}\left(\frac{b_s}{2}\right)^n,  \label{Bess0}\\
&2 \frac{\ju}{\aal}=\mathrm{e}^{-b_s/2}\sum_{n=0}^{+\infty} \frac{L_n^{(1)}\left( \frac{\mu_s B_0}{\Tpea}\right)}{(n+1) !}\left(\frac{b_s}{2}\right)^n, \label{Bess1}
\end{align}
where $L_n^{(1)}$ is a generalized Laguerre polynomial.
 Making use of the Fourier representations (\ref{fmnFourier}) , (\ref{fourier1}) and (\ref{fourier2}) for for $f_{{mn}_s}$, $\wphi$ and $\wbpar$, respectively, one can deduce from Eq. (\ref{closk}) the relation
 \beq  \label{qsclos}
 f_{{mn}_s}(\bx ,t)=-\delta_{m0}\left( G_{1ns} \frac{\qa}{\Tpa} \wphi+2 \thea G_{2ns} \frac{\wbpar}{B_0}\right),
 \eeq
 or, equivalently,
 \beq  \label{closmn}
 f_{{mn}_s}(\bx ,t)=-\delta_{m0}\thea \left(\sgn(\qa)G_{1ns}\frac{\phi}{\taupa}+2G_{2ns} \bpar\right),
 \eeq
 where we also made use of the normalization (\ref{norm3}) for $\phi$ and $\bpar$.  The operators $G_{1ns}$ and $G_{2ns}$ are defined, consistently with the definition of $G_{10s}$ and $G_{20s}$ given in Sec. \ref{sec:gyrofluid}, by $G_{1ns} f (\bx ,t)=\sum_{\bk \in \mathscr{D}} G_{1n} (b_s) f_{\bk}(t) \exp(i \bk \cdot \bx)$ and $G_{2ns} f (\bx ,t)=\sum_{\bk \in \mathscr{D}} G_{2n} (b_s) f_{\bk}(t) \exp(i \bk \cdot \bx)$, for a function $f$ and $n\geq 0$ (in the specific case of the linear dispersion relation, the dependence on $t$ is provided by the factor $\mathrm{e}^{-i \omega t}$, but when the relations (\ref{closmn})  are used as closures for the nonlinear models, the dependence on $t$ is of course left arbitrary).
 
 The relations (\ref{closmn}) descending from the quasi-static assumption, are adopted as closure relations in GF6 and GF6L for all the moments involved in the model, except for $N_s$, $M_s$, $\Tps$, which are derived  by solving the evolution equations (\ref{conta})-(\ref{tempa}). The parallel heat flux fluctuations $\qps$, on the other hand, are determined, as already mentioned,  again by a quasi-static closure  for GF6 (Eq. (\ref{closgf6}) which follows from Eq. (\ref{closmn}) for $m=3, n=0$) or by the Landau closure (\ref{landau}) for GF6L. The closure (\ref{closgf4}) adopted for GF4, is again a quasi-static closure following from Eq. (\ref{closmn}) when $m=2$ and $n=0$.
  
 \section{Derivation of the model equations} \label{sec:appb}
 
 The gyrofluid system (\ref{conta})-(\ref{ampea}) descends from the parent gyrokinetic system (\ref{gyr})-(\ref{amppedim}) upon applying to the perturbations of the distribution functions the expansion (\ref{expf}). In order to obtain a closed system with a finite number of equations, such expansion is constrained in the following way. The moments $f_{00_s}, f_{10_s}$ and $f_{20_s}$ (or, equivalently, by virtue of Eqs. (\ref{momf0})-(\ref{momf}), the gyrofluid densities, parallel velocities and temperatures $N_s$, $U_s$ and $\Tps$), for each species $s$, get determined by evolution equations obtained by making the product of all the terms of the gyrokinetic equation (\ref{gyr}) with the zero, first and second order Hermite polynomial in the variable $\vpar/ \vtpa$ and integrating over the velocity volume element $\dwa$. For GF6L, the parallel heat flux $\qps$ gets determined by the relation (\ref{landau}). All the other moments, on the other hand,  are assumed to be given by the relations (\ref{qsclos}), or, in normalized form, by Eq. (\ref{closmn}), obtained from the linear theory in the quasi-static limit. With this prescription, the expansion (\ref{expf}) becomes

 \begin{align}
  &\dfa (\bx,\vpar,\mua,t)  =\calfa (\vpar ,\mua)\left(\frac{\wns}{n_0}(\bx ,t) +\frac{\vpar}{\vtpa}\frac{\wus}{\vtpa} (\bx ,t)  \right.  \nno \\
  &\left. + \frac{1}{2}\left(\frac{\vpar^2}{\vtpa^2}-1\right)\frac{\wtps}{\Tpa} (\bx ,t) + \frac{1}{6} \left( \frac{\vpar^3}{\vtpa^3}- 3 \frac{\vpar}{\vtpa}\right)\frac{\qpars}{n_0 \Tpa \vtpa}  (\bx ,t) \right. \nno \\
  & \left. -\sum_{n =1}^{+\infty} L_n\left(\frac{\mua B_0}{\Tpea}\right) \left( G_{1ns}  \frac{\qa}{\Tpa} \wphi (\bx ,t)+2 \thea G_{2ns}   \frac{\wbpar}{B_0} (\bx ,t)\right)\right),   \label{expclos}
 \end{align}
 where we made use of the fact that $H_0 (\vpar / \vtpa)=1$, $H_1(\vpar / \vtpa)=\vpar / \vtpa$, $H_2 (\vpar / \vtpa)=\vpar^2/\vtpa^2-1$ and $H_3(\vpar / \vtpa)=\vpar^3 / \vtpa^3 - 3 \vpar / \vtpa$.
 
 Inserting the expansion (\ref{expclos}) into Eqs. (\ref{defg}) and  (\ref{gyr}), and integrating over $\dwa$ one obtains
 \begin{align}
 &\frac{\pa}{\pa t} \frac{\wns}{n_0}+\frac{c}{B_0}\left(\left[ G_{10s}  \wphi , \frac{\wns}{n_0}\right]-\sum_{n=1}^{+\infty} \left[ G_{1ns}  \wphi ,  \frac{\qa}{\Tpa} G_{1ns} \wphi + 2 \thea G_{2ns}   \frac{\wbpar}{B_0}\right] \right. \nno \\
 &\left. -\sum_{n=1}^{+\infty} \left[ 2 \frac{\Tpea}{\qa}G_{2ns}  \frac{\wbpar}{B_0} , \frac{\qa}{\Tpa}G_{1ns}  \wphi + 2 \thea G_{2ns}   \frac{\wbpar}{B_0}\right] +\frac{\Tpea}{\qa}\left[2 G_{20s} \frac{\wbpar}{B_0} , \frac{\wns}{n_0}\right]\right.  \label{n1} \\
 & \left. -\left[ G_{1ns} \wapar , \frac{\wus}{c}\right]\right)+\frac{\pa \wus}{\pa z}=0, \nno
 \end{align}
 where we made use of the definitions (\ref{gyroop1}) and (\ref{gyroop2}) and of the orthogonality of Hermite polynomials. We remark, at this point, that the sum of the last term in the first line of Eq. (\ref{n1}) with the first term on the second line of Eq. (\ref{n1}) yields zero, because of the antisymmetry of the canonical bracket $[ \, , \, ]$.  We thus conclude that  the quasi-static closure (\ref{qsclos}) has the remarkable property of annihilating, in the continuity equation, all the contributions associated with the moments $f_{0n_s}$, for $n \geq 1$. By virtue of this cancellation, from Eq. (\ref{n1}) we obtain
 \begin{align}
 &\frac{\pa}{\pa t} \frac{\wns}{n_0}+\frac{c}{B_0}\left[ G_{10s}  \wphi +\frac{\Tpea}{\qa} 2 G_{20s} \frac{\wbpar}{B_0} , \frac{\wns}{n_0}\right] -\frac{1}{B_0}\left[ G_{10s} \wapar , \wus\right]+\frac{\pa \wus}{\pa z}=0. \label{n2}
 \end{align}
 Applying to Eq. (\ref{n2}) the normalization (\ref{norm1})-(\ref{norm4}), one obtains Eq. (\ref{conta}).
 
 In order to derive Eq. (\ref{ua}) we point out first that, with the help of the identities
 \begin{align}
& \jo=\mathrm{e}^{-b_s/2}\sum_{n=0}^{+\infty} \frac{L_n\left( \frac{\mu_s B_0}{\Tpea}\right)}{n !}\left(\frac{b_s}{2}\right)^n  ,  \label{Bessel0}\\
&2 \frac{\ju}{\aal}=\mathrm{e}^{-b_s/2}\sum_{n=0}^{+\infty} \frac{L_n^{(1)}\left( \frac{\mu_s B_0}{\Tpea}\right)}{(n+1) !}\left(\frac{b_s}{2}\right)^n , \label{Bessel1}\\
& \int_0^{+\infty} dx \, \mathrm{e}^{-x} L_n(x) L_m(x) =\delta_{mn},  \label{ortlag} \\
&L_m^{(1)} (x)=-\frac{d}{dx}L_{m+1}(x)=-\frac{m+1}{x}L_{m+1}(x)+\frac{m+1}{x}L_m (x),
\end{align}
we obtain the relation (see also \citet{Bri92})
\begin{align}
&\frac{c}{ B_0}\int \dwa \, \calfa \left[ \mathcal{J}_{0s} \wphi , \frac{\qa}{\Tpa} \frac{\vpar^2}{\vtpa c} \mathcal{J}_{0s} \wapar \right]\\
&=n_0 \frac{\vtpa}{ B_0}\sum_{\bk , \bk ' \in \scrd}  \int d\left(\frac{\mua B_0}{\Tpea}\right) \, \mathrm{e}^{-\mua B_0 / \Tpea} \nno \\
&\times  \left[ \frac{\qa}{\Tpa} J_0\left(\frac{k_\perp}{\omega_{cs}}\sqrt{\frac{2 \mua B_0}{m_s}}\right)\wphi , J_0\left(\frac{k_{\perp}^{'} }{\omega_{cs}}\sqrt{\frac{2 \mua B_0}{m_s}}\right)\frac{\qa}{\Tpa}\wapar \right] \mathrm{e}^{i (\bk + \bk ')\cdot \bx}\\
&=n_0 \frac{\vtpa}{ B_0}\sum_{\bk , \bk ' \in \scrd} \sum_{m,n=0}^{+\infty} \int d\left(\frac{\mua B_0}{\Tpea}\right) \,  \mathrm{e}^{-\mua B_0 / \Tpea} \nno \\
& \times \left[ \frac{\qa}{\Tpa}\mathrm{e}^{-b_s/2} \frac{L_m\left( \frac{\mu_s B_0}{\Tpea}\right)}{m !}\left(\frac{b_s}{2}\right)^m \wphi,
\mathrm{e}^{-b_s^{'}/2} \frac{L_n\left( \frac{\mu_s B_0}{\Tpea}\right)}{n !}\left(\frac{b_s^{'}}{2}\right)^n \wapar\right] \mathrm{e}^{i (\bk + \bk ')\cdot \bx} \nno \\
&= n_0 \frac{\vtpa}{ B_0}  \sum_{n=0}^{+\infty} \left[ \frac{\qa}{\Tpa} G_{1ns}  \wphi , G_{1ns}  \wapar \right],  \label{intjojo}
\end{align}
and, by an analogous procedure, the relation 
\begin{align}
&\frac{c}{B_0}\int \dwa \, \calfa \left[  \frac{2 \mu_s B_0}{\qa} \mathcal{J}_{1s} \frac{\wbpar}{B_0} , \frac{\qa}{\Tpa} \frac{\vpar^2}{\vtpa c} \mathcal{J}_{0s} \wapar \right]   \nno \\
&=n_0 \thea \frac{\vtpa}{B_0}  \sum_{n=0}^{+\infty} \left[ 2 G_{2ns}  \frac{\wbpar}{B_0} , G_{1ns}  \wapar \right].  \label{intjoj1}
\end{align}
Upon multiplying Eq. (\ref{gyr}) by $(1/n_0) \vpar / \vtpa$ and integrating over $\dwa$ one obtains, adopting the expansion (\ref{expclos}) as well as the relations (\ref{gyroop1}), (\ref{gyroop2}), (\ref{intjojo}) and (\ref{intjoj1}), the following equation
\begin{align}
& \frac{\pa}{\pa t} \left( \frac{\wus}{\vtpa} + \frac{\qa}{\Tpa}\frac{\vtpa}{c} \gamus  \wapar\right)+\frac{c}{B_0}\left[ \gamus  \wphi +\frac{\Tpea}{\qa} 2 \gamds \frac{\wbpar}{B_0} , \frac{\wus}{\vtpa}\right] \nno \\
&+ \frac{\vtpa}{B_0} \sum_{n=0}^{+\infty} \left[ \frac{\qa}{\Tpa} G_{1ns}  \wphi +2 \thea  G_{2ns}  \frac{\wbpar}{B_0} , G_{1ns}  \wapar \right] -\frac{\vtpa}{B_0} \left[ \gamus \wapar , \frac{\wns}{n_0}+\frac{\wtps}{\Tpa}\right] \nno \\
& + \frac{\vtpa}{B_0} \sum_{n=1}^{+\infty} \left[ G_{1ns}  \wapar , \frac{\qa}{\Tpa}  G_{1ns}  \wphi + 2 \thea G_{2ns}  \frac{\wbpar}{B_0} \right] \label{u1} \\
&+\vtpa \frac{\pa}{\pa z} \left( \frac{\wns}{n_0} +  \frac{\qa}{\Tpa}  \gamus  \wphi + 2 \thea \gamds \frac{\wbpar}{B_0} +\frac{\wtps}{\Tpa}\right)=0. \nno
\end{align}
Also in this case, the quasi-static closure leads to a remarkable cancellation. Indeed, among the nonlinear terms involving only electromagnetic quantities (i.e. $\wphi$, $\wapar$ and $\wbpar$) all those containing gyroaverage operators $G_{1ns}$ and $G_{2ns}$, with $n \geq 1$, vanish. As a result, Eq. (\ref{u1}) reduces to
\begin{align}
& \frac{\pa}{\pa t} \left( \frac{\wus}{\vtpa} + \frac{\qa}{\Tpa}\frac{\vtpa}{c} \gamus  \wapar\right)+\frac{c}{B_0}\left[ \gamus  \wphi +\frac{\Tpea}{\qa} 2 \gamds \frac{\wbpar}{B_0} , \frac{\wus}{\vtpa}\right] \nno \\
& + \frac{\vtpa}{B_0} \left[ \frac{\qa}{\Tpa} \gamus  \wphi + 2 \thea \gamds \frac{\wbpar}{B_0} , \gamus  \wapar \right] - \frac{\vtpa}{B_0} \left[ \gamus \wapar , \frac{\wns}{n_0} +\frac{\wtps}{\Tpa}\right] \label{u2} \\
& + \vtpa \frac{\pa}{\pa z} \left( \frac{\wns}{n_0} +  \frac{\qa}{\Tpa}  \gamus  \wphi + 2 \thea \gamds \frac{\wbpar}{B_0} +\frac{\wtps}{\Tpa}\right)=0. \nno
\end{align}
Equation (\ref{ua}) is then obtained from Eq. (\ref{u2}) after applying the normalization (\ref{norm1})-(\ref{norm4}).

Equation (\ref{tempa}) is obtained upon multiplying Eq. (\ref{gyr}) by $(1/n_0) (\vpar^2/ \vtpa^2 -1)$ and integrating over $\dwa$. Making use of the expansion (\ref{expclos}) and of the orthogonality of Hermite polynomials one obtains
\begin{align}
&\frac{\pa}{\pa t}\frac{\wtps}{\Tpa}+\frac{c}{B_0}\left( \left[ G_{10s}  \wphi , \frac{\wtps}{\Tpa}\right]+\frac{\Tpea}{\qa}\left[2 G_{20s} \frac{\wbpar}{B_0} , \frac{\wtps}{\Tpa}\right] \right. \nno\\
&\left. -2  \left[ G_{10s} \wapar , \frac{\wus}{c}\right]- \left[ G_{10s}\wapar , \frac{\qpars}{n_0 \Tpa c}\right]\right) \nno \\
&+ \vtpa \frac{\pa}{\pa z}\left(2 \frac{\wus}{\vtpa}+\frac{\qpars}{n_0 \Tpa \vtpa}  \right)=0.  \label{tempdim}
\end{align}
Adopting the normalization (\ref{norm1})-(\ref{norm4}), Eq. (\ref{tempa}) follows from Eq. (\ref{tempdim}).

Equations (\ref{qna}), (\ref{ampa}) and (\ref{ampea}) follow from Eq. (\ref{qndim}), (\ref{ampdim}) and (\ref{amppedim}), respectively, upon inserting the expansion (\ref{expclos}), evaluating the integrals and applying the normalization (\ref{norm1})-(\ref{norm4}). With regard to the evaluation of the integrals and the derivation of the equations in the form (\ref{qna}), (\ref{ampa}) and (\ref{ampea}), we remark that, in addition to Eqs. (\ref{gyroop1})-(\ref{gyroop2}), the following relations  are of use:
\begin{align}
&\frac{1}{n_0}\int \dwa \, \calfa J_0^2 (\aal)=\gamz (b_s)=\sumz G_{1n}^2 (b_s), \\
& \frac{1}{n_0}\int \dwa \, \calfa \frac{2 \mu_s B_0}{\Tpea} \jo \frac{\ju}{\aal}=\gamz (b_s) - \gammu (b_s)=G_{10}^2 (b_s) +2 \sum_{n=1}^{+\infty} G_{1n} (b_s) G_{2n} (b_s), \\
&\frac{1}{n_0}\int \dwa \, \calfa \left( \frac{2 \mu_s B_0}{\Tpea}  \frac{\ju}{\aal} \right)^2=2 (\gamz (b_s) - \gammu (b_s))=G_{10}^2 (b_s)+4 \sum_{n=1}^{+\infty} G_{2n}^2 (b_s).
\end{align}

\section{Field-swelling instabilities} \label{App:swelling}

\subsection{Dispersion relation at the MHD scales}

In this Appendix, we first provide a simple derivation of the dispersion relation for fast and slow modes at MHD scales in the presence of temperature anisotropy, starting from the kinetic-MHD equations (see Eqs. (37), (38b), (44 a,b,c) (46)-(48) from \citet{K83}, or equivalently Eqs. (1)-(8) from \citet{Snyder97}). 
	
The transverse velocity can be decomposed into compressible and solenoidal  parts by writing
\begin{equation}
u_{\perp } = - \nabla_\perp \chi_{c  } + \nabla_\perp \times (\chi_{ s } {\widehat z}).
\end{equation}
	
One immediately gets (see e.g. Eqs. (48),(52) and (56) of \citet{PS06} where FLR corrections are neglected) 
\begin{eqnarray}
&&-\partial_t\partial_{xx}\chi_{c}+\partial_{xx} \Big
(\frac{p_\perp}{{\rho_0}}+c_A^2 \frac{B_\|}{B_0}\big ) + \big (c_A^2+{\frac{p_{0_\perp}-p_{0_\|}}{\rho_0}}\big)\partial_{zz}\frac{B_\|}{B_0}=0, \label{fluid:chic}\\
&& \partial_t \frac{B_\|}{B_0}-\partial_{xx} \chi_c =0 , \label{fluid:B_z}
\end{eqnarray}
where $p_\perp$ and $p_\|$ denote the total (ion plus electron) perpendicular and parallel pressure fluctuations, 
with the subscript $0$ referring to the equilibrium values. Furthermore, $\rho_0$ denotes the equilibrium plasma density. 
	
In Eq. (\ref{fluid:chic}), the perpendicular pressure fluctuations are given by the drift-kinetic theory, as found e.g. in Eq. (27) of \citet{Snyder97}, in the form
\begin{align}
\frac{p_{\perp r}}{\rho_0}=\frac{T_{0_{\perp r}}}{m_i}\left ( 2(1-\Theta_r R(\zeta_r))\frac{B_\|}{B_0} -R(\zeta_r)\frac{{q}_r\psi}{T_{\| r}}\right ),
\end{align}
together with
\begin{equation}
\frac{n_r}{n_0}=(1-\Theta_r R(\zeta_r))\frac{B_\|}{B_0}-R(\zeta_r)\frac{{q}_r\psi}{T_{\| r}},
\end{equation}
where ${q}_r=e$ for ions and $-e$ for electrons, respectively. The potential $\psi$ is defined in terms of the parallel electric field by $E_z = -\partial_z \psi$.
These expressions for the perpendicular pressure and density perturbations identify with the large-scale limit of formulas given in Appendix B of \citet{PS07}.
Quasineutrality requires the equality  of the electron ($n_e$)  and ion ($n_i$)  number-density fluctuations, which prescribes 
\begin{equation}
\frac{e\psi}{\Tpae}=\tau_\|\frac{\Theta_e R(\zeta_e)-\Theta_iR(\zeta_i)}{\tau_\| R(\zeta_e)+R(\zeta_i)}\frac{B_\|}{B_0}.
\end{equation}
Plugging the expressions for the pressures in Eqs. (\ref{fluid:chic})-(\ref{fluid:B_z}) thus leads to
\begin{align}
\frac{\omega^2}{k^2 c^2_A}=&1-\frac{k^2_z}{k^2}\frac{\beta_{\| e}}{2}(1-\Theta_e+\tau_\|(1-\Theta_i))\nonumber\\
&+\frac{k^2_\perp}{k^2}\frac{\beta_{\perp e}}{2}\Big [2(1-\Theta_eR(\zeta_e))+\taupi (2-\Theta_eR(\zeta_e)-\Theta_iR(\zeta_i))\nonumber \\
&+\tau_\|(1+\taupi)R(\zeta_e)\Big ( \frac{\Theta_eR(\zeta_e)-\Theta_iR(\zeta_i)}{\tau_\| R(\zeta_e)+R(\zeta_i)}\Big ) \Big]. \label{disp}
\end{align}
We note that $\displaystyle{\zeta^2_i=\frac{\omega^2}{2\tau_\| \omega^2_s}}$ with $\omega_s= k_z c_{s_\|}$ where  $\displaystyle{c_{s_\|}=\sqrt{\frac{\Tpae}{m_i}}}$ is the sound speed based on the parallel electron temperature.
	
\subsection{Link with the fluid theory}
	
In order to make a link with fluid theory as performed in Basu \& Coppi (1984), we note that keeping all the terms in the ion density and parallel velocity equations (in particular the time derivatives), is equivalent to expanding $R(\zeta_i)$ for $\zeta_i$ large, i.e. in the adiabatic limit where $R(\zeta_i)\approx -1/(2\zeta_i^2)$. It thus follows that the ions cannot be hot, at least outside an angular boundary layer near the transverse direction. 
	
In addition to adiabatic ions, let us also assume a quasi-static limit for the electrons, which leads to assume $\zeta_e \to 0$ and thus $R(\zeta_e)\approx 1$. Equation (\ref{disp}) then rewrites
\begin{align}
\frac{\omega^2}{k^2 c^2_A}=&1-\frac{k^2_z}{k^2}\frac{\beta_{\| e}}{2}(1-\Theta_e+\tau_\|(1-\Theta_i))\nonumber\\
&+\frac{k^2_\perp}{k^2}\frac{\beta_{\perp e}}{2}\Big [2(1-\Theta_e)+\taupi \left(2-\Theta_e+\frac{\Theta_i}{2\zeta_i^2}\right) \nonumber\\
&+\tau_\|(1+\taupi)\Big ( \frac{\Theta_e+\Theta_i/(2\zeta_i^2)}{\tau_\| -1/(2\zeta_i^2)}\Big ) \Big]. \label{disp2}
\end{align}
For cold ions ($\tau_\|=\taupi=0$), we get  
	\begin{align}
		&\left ( \frac{\omega^2}{\omega_s^2}-1\right)\left [ 1-\frac{\omega^2}{k^2 v^2_A} -\frac{k^2_z}{2k^2}(\beta_{\| e} -\beta_{\perp e}) 
		+\frac{k^2_\perp}{k^2}\Theta_e (\beta_{\| e}-\beta_{\perp e})\right]\nonumber\\ 
		&+\frac{k^2_\perp}{2k^2}\Theta_e\beta_{\perp e}\frac{\omega^2}{\omega_s^2}=0, \label{disp3}
	\end{align}
	which also rewrites in the form of Eq. (44) of \citet{Bas84}
	\begin{align}
		&\left ( \frac{\omega^2}{\omega_s^2}-1\right)\left [ 1-\frac{\omega^2}{k^2 v^2_A} 
		+\frac{k^2_\perp}{2k^2}\Theta_e (2\beta_{\| e}-\beta_{\perp e})-\frac{k^2_z}{2k^2}(\beta_{\| e} -\beta_{\perp e}) \right]\nonumber\\
		&+\frac{k^2_\perp}{2k^2}\Theta_e\beta_{\perp e}=0. \label{disp4}
	\end{align}
	
	The slow mode is obtained when $\omega \sim \omega_s$, while the fast mode corresponds to $\omega\gg \omega_s$.
	
	\bigskip
	\subsection{The field swelling instabilities}
	\subsubsection{Slow mode}
		As discussed in \citet{Bas84}, it follows from Eq. (\ref{disp4}) that the slow mode becomes unstable when
		\begin{equation}
			1<\Theta_e \frac{\beta_{\perp e}}{1+\beta_{\perp e}}<2.
		\end{equation}
		As $\Theta_e$ is increased from 1, the phase velocity decreases and becomes zero at $1+1/\beta_{\perp e}$. As $\Theta_e$ is further increased, a pair of purely imaginary complex conjugate roots appears, leading to the so-called slow-mode swelling instability.
		\bigskip
\subsubsection{ Fast mode}
		According to the theory of \citet{Bas84} that assumes $\zeta_e$ very small, the fast mode becomes unstable when 
		\begin{equation}
			\Theta_e >\frac{2(1+\beta_{\perp e})}{\beta_{\perp e}}.
		\end{equation}
The validity conditions require in particular that $(k_z/k_\perp)^2>(m_e/m_i)(1/\beta_{\| e})$ in order to ensure that $\zeta_e$ is small enough.

\subsection{Instability growth rate}
	
\begin{figure}
\centerline{
\includegraphics[width=0.31\textwidth]{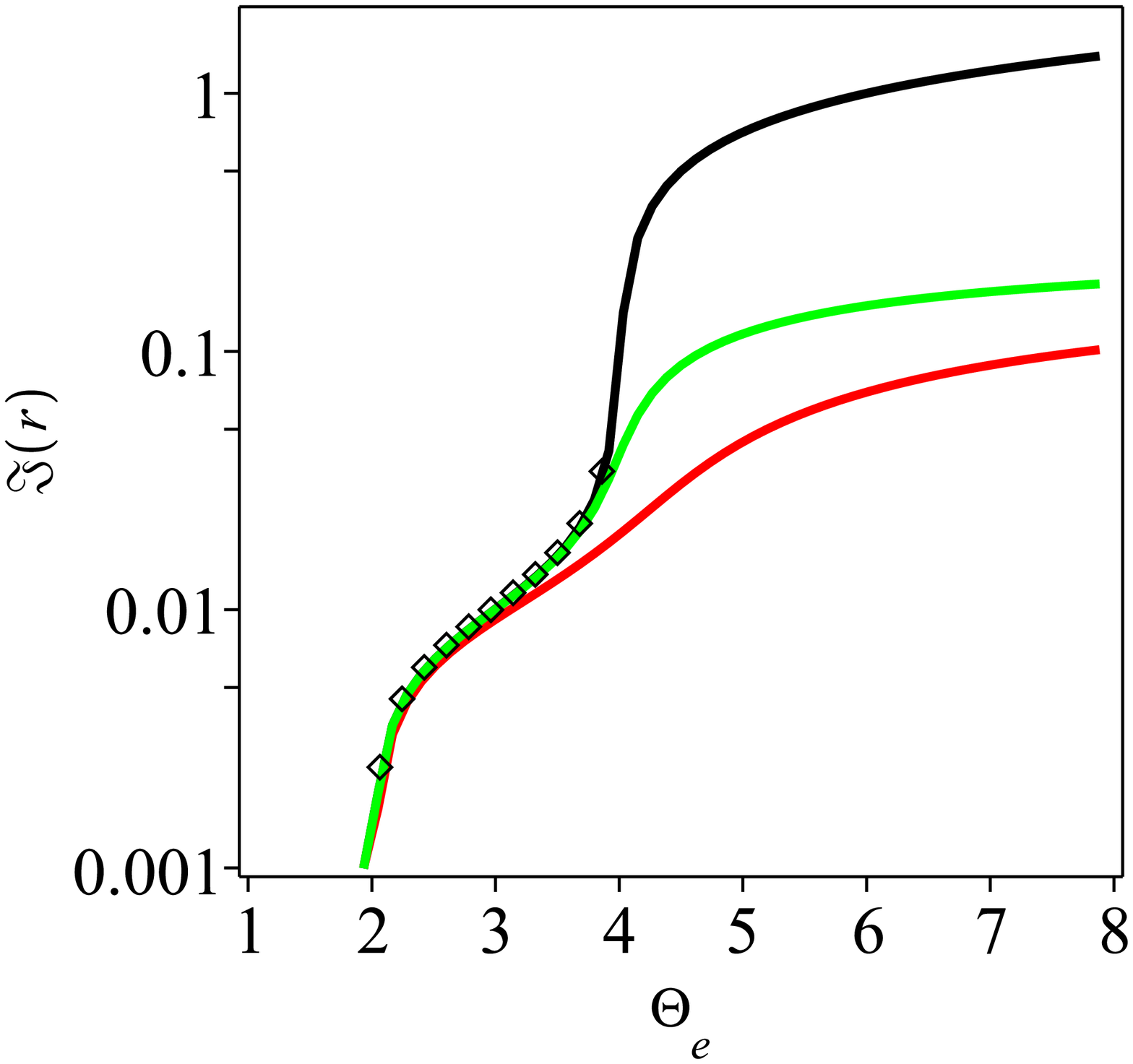}
\includegraphics[width=0.31\textwidth]{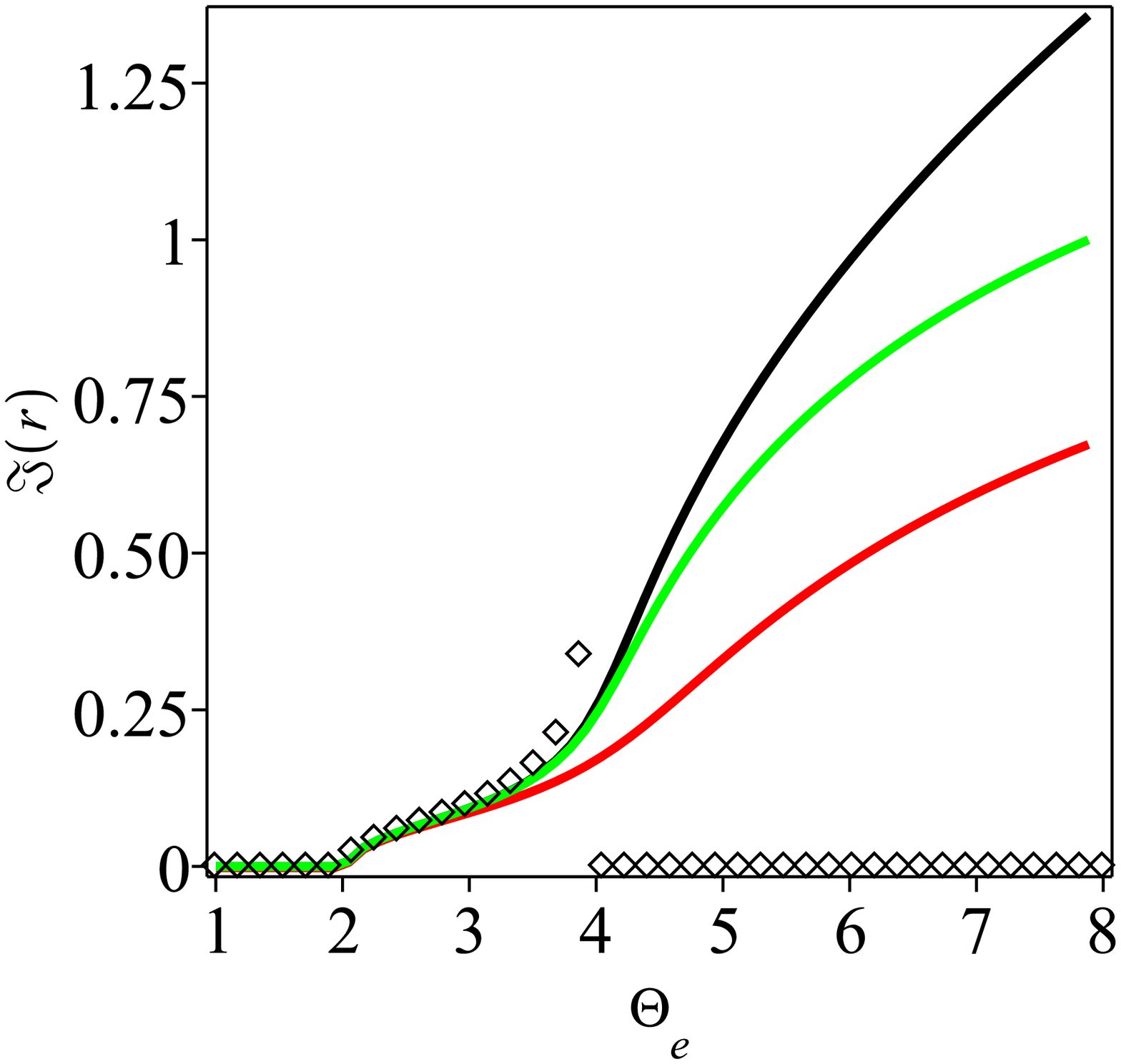}
\includegraphics[width=0.31\textwidth]{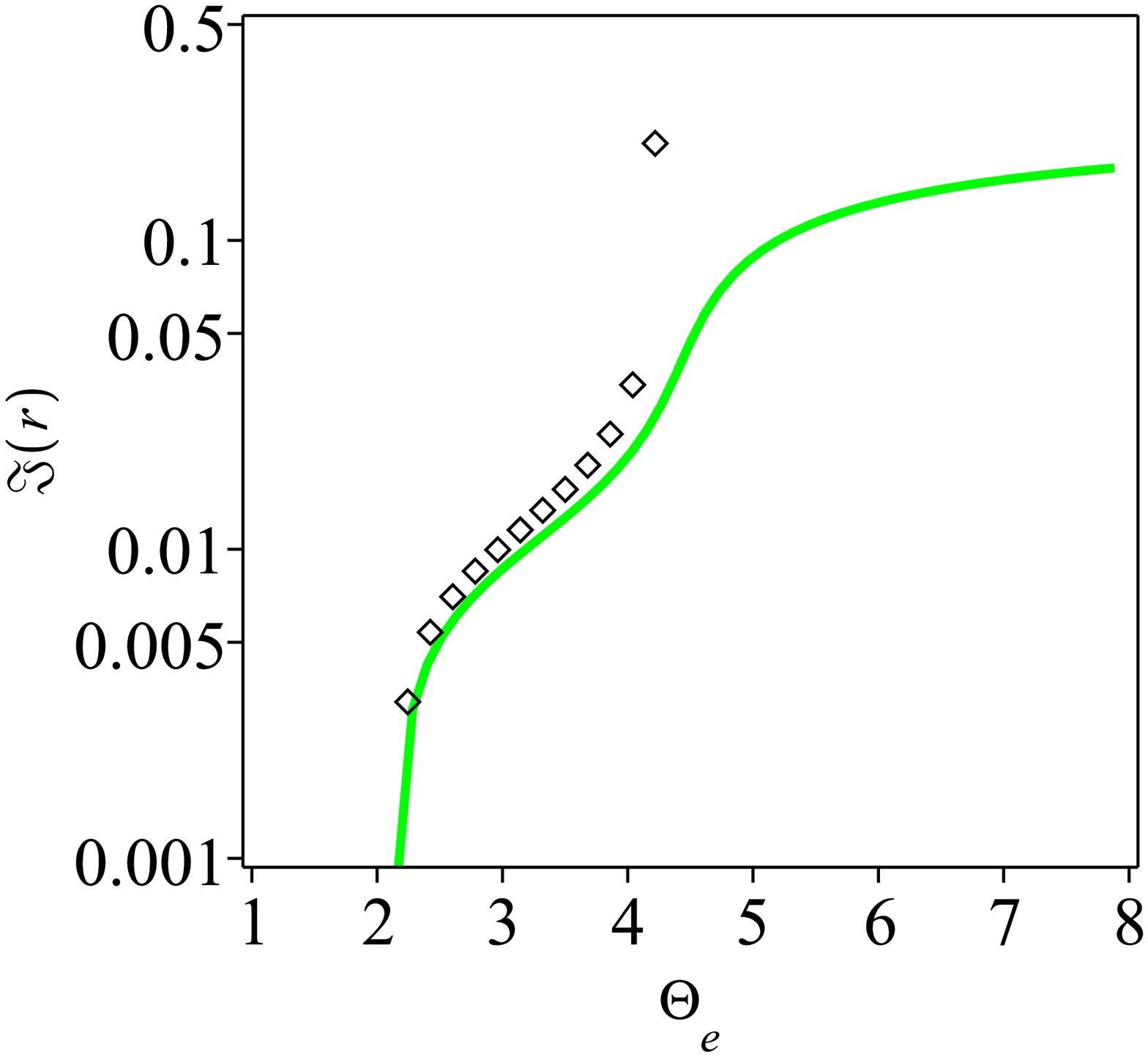}}
\caption{Imaginary part of $r=\omega/kc_A$ for  the unstable mode, solution of Eq. (\ref{disp}), for propagation angles $\alpha=89^\circ$ (left) and $80^\circ$ (middle) as a function of the perpendicular to parallel electron temperature anisotropy $\Theta_e$ in the case  $\beta_{\perp e}=1$ and cold ion temperatures (leading to take $R(\zeta_i) = -1/(2\zeta_i^2)$)
for various approximations of $R(\zeta_e)$: BC84 (black solid line) uses   $R(\zeta_e) = 1$, DK (red solid line) uses  $R_{20}(\zeta_e)$ and  DKNL (green solid line) uses ${ R_{20}}(\zeta_e)$ (no electron Landau damping).  Superimposed diamond symbols refer to predictions of GF4 model taken in the large-scale limit. 
Right panel corresponds to DKNL and GF4 models with $\alpha=89^\circ$ in the case  $\tau_{\perp i}=1$.}
\label{fig:swelling-vs-ae}
\end{figure}

Le us now  consider the full dispersion relation, given by Eq. (\ref{disp}), with a general electron response function.
The instabilities are illustrated in Fig. \ref{fig:swelling-vs-ae} which displays, as a function of the perpendicular to parallel electron temperature anisotropy $\Theta_e$, the imaginary part of $r=\omega/kc_A$ for  the unstable mode, solution of Eq. (\ref{disp}) with $\beta_{\perp e}=1$ and cold ion temperatures, for propagation angle $\alpha = 89^{\circ}$ (left) and $\alpha= 80^{\circ}$ (middle), when using  four different approximations for the plasma response functions. The first one, referred to as BC84 (black solid line) uses  $R(\zeta_i)= -1/(2\zeta_i^2)$ and $R(\zeta_e)=1$ and corresponds to the asymptotic  solution of \citet{Bas84}, the second one, denoted DK (red solid line) uses for the electrons the Pad\'e approximant  $R_{20}(\zeta_e)=1/(1-2\zeta_e^2-i\pi^{1/2}\zeta_e)$ (the use of $R_{42}$ leads to almost identical results) and for the ions the same large-$\zeta_i$ limit as in the   BC84 model, since the ions are cold. The third model, called DKNL (green solid line) differs from the previous one by the fact that electron Landau damping is suppressed (using $R_{21}(\zeta_e)=1/(1-2\zeta_e^2$)). 
Superimposed diamond symbols refer to the prediction of  GF4, taken in the large-scale limit. It is legitimate to consider this model, derived in the limit $\zeta_i\to 0$, in the regime of cold ions, because in this case the dynamics is insensitive to the closure assumption. 
The right panel corresponds to the case with $\taupi=1$ for $\alpha=89^\circ$. It shows that the GF4 model is still approximately valid even in a situation where $\zeta_i$ is not small. 
	
\begin{figure}
		\centerline{
			\includegraphics[width=0.48\textwidth]{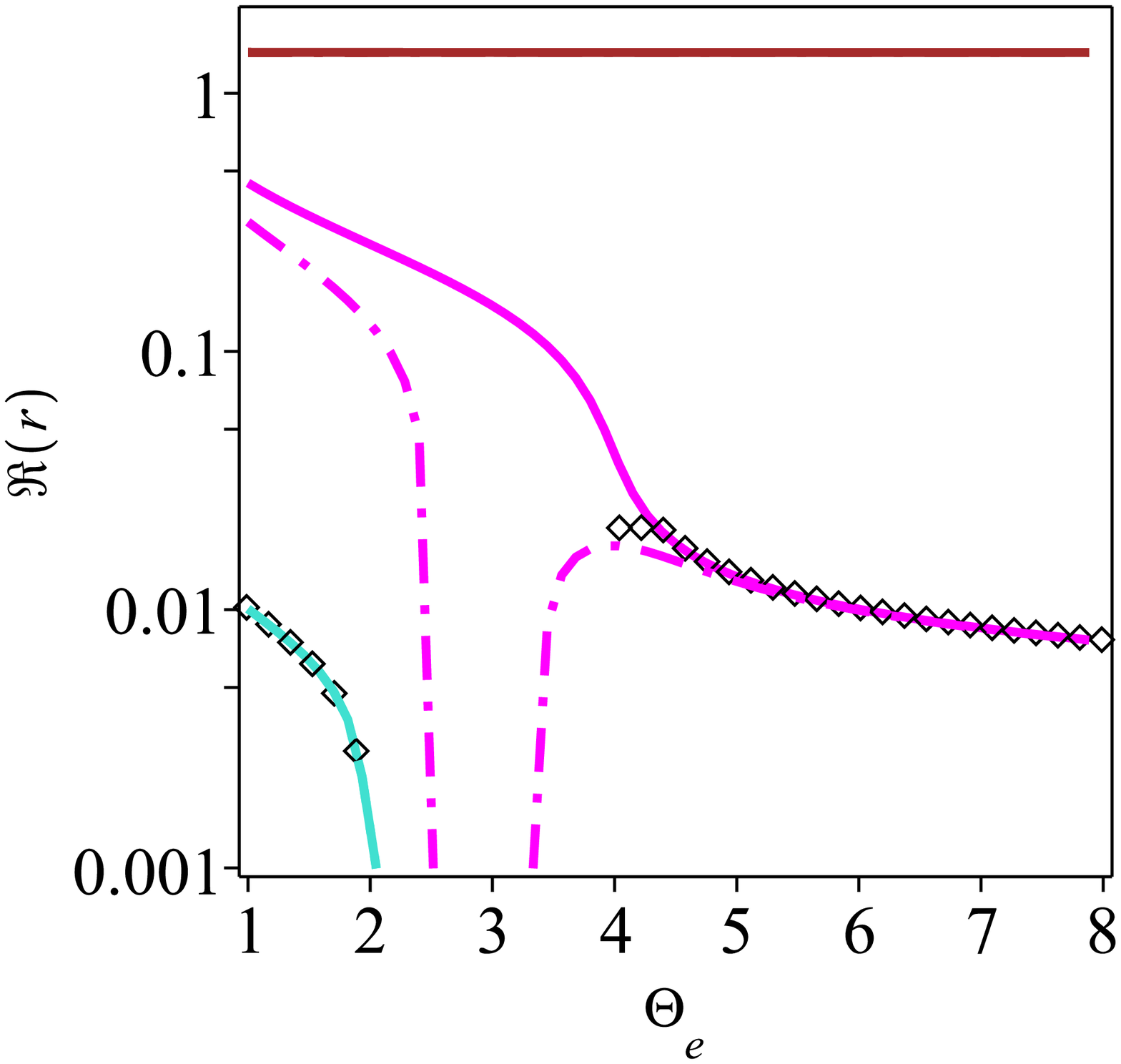}
			\includegraphics[width=0.48\textwidth]{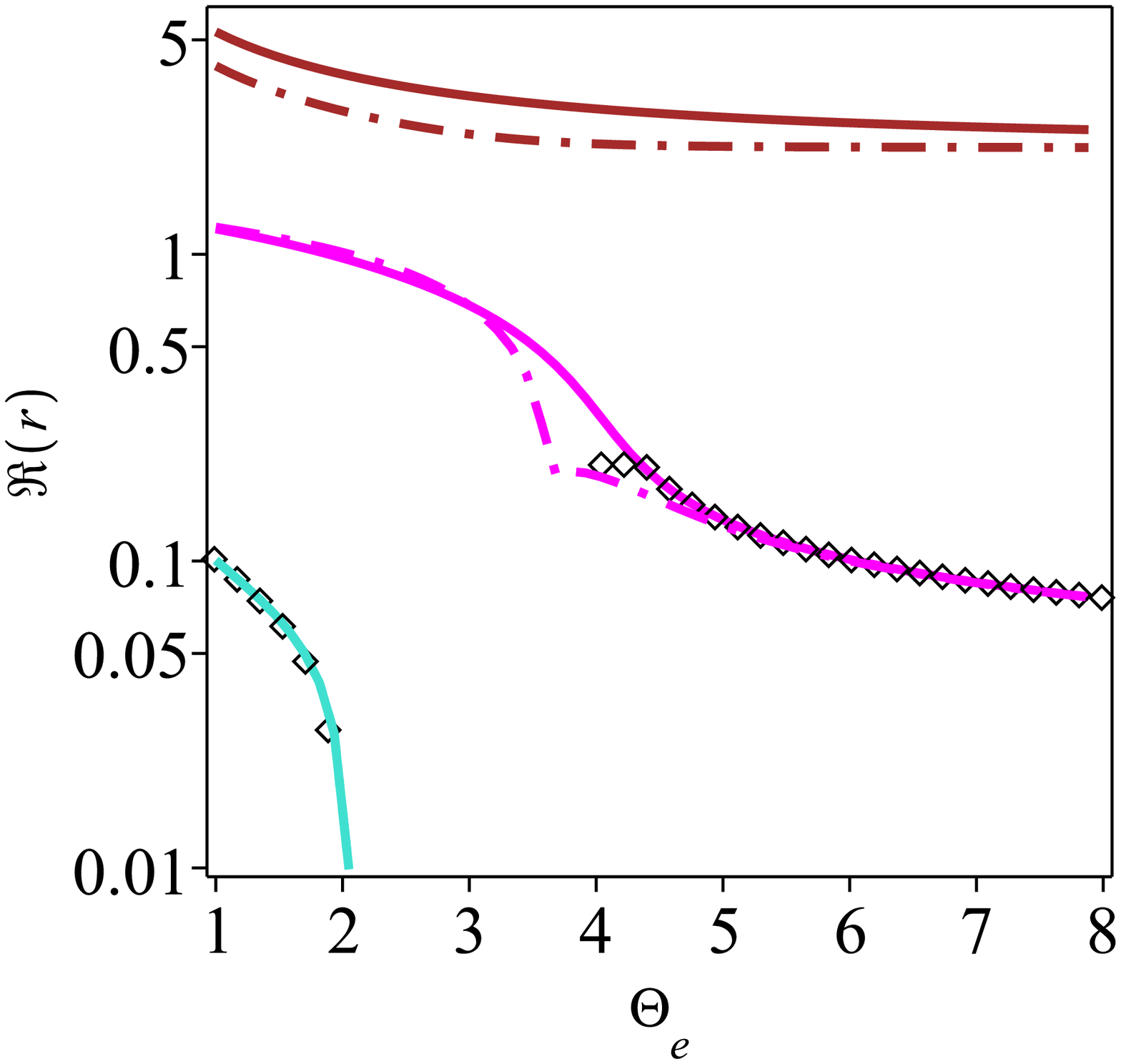}}
		\caption{Positive real part of the three roots (denoted min (turquoise), int (magenta) and max(brown) in increasing order of magnitude for DKNL  (solid lines) and DK  (dash-dotted lines), together with predictions of  GF4  (diamond symbols), versus $\Theta_e$, for  $\alpha = 89^\circ$ (left) and  $\alpha = 80^\circ$ (right), in the case of cold ions.}
		\label{fig:minmaxint}
	\end{figure}

Figure \ref{fig:minmaxint} displays, as a function of  $\Theta_e$, the positive real part of the three roots of Eq. (\ref{disp}) (referred to as {\it min}, {\it int} and {\it max} in increasing order of magnitude, displayed in turquoise, magenta and brown colors respectively), in the case of DKNL model (solid lines) or for the DK model (dash-dotted lines), with the predictions of the GF4 model superimposed as diamond symbols, again for $89^\circ$ (left) and $80^\circ$ (right) propagation angles  and  cold ions.
For the $89^\circ$  angle (which falls outside the range of admissible angles for the BC84 model), the destabilization of the slow mode for $\Theta_e>2$ is almost similar for the four models. Within the BC84 model, the fast mode becomes unstable for $\Theta_e>4$: the real part of the associated root vanishes (not shown) and the instability growth rate (black solid line of the  left panel of Fig. \ref{fig:swelling-vs-ae}) increases rapidly. The slow mode reappears when $\Theta_e>4$. The DKNL model displays a very different behavior, whereby the fast mode (brown solid line in the left panel of Fig. \ref{fig:minmaxint}) remains almost unchanged for the whole range of values of $\Theta_e$. The slow mode (turquoise solid line in the same panel) disappears for $\Theta_e>2$ and reappears on the intermediate branch (magenta solid line of the left panel of Fig. \ref{fig:minmaxint}) for $\Theta_e>4$.  The intermediate root for $\Theta_e <4$ is one of the many plasma modes that coexist with the usual slow and fast modes of fluid theory; it is here the only extra mode for the present choice of the Pad\'e approximant ($R_{21}(\zeta_e)$, leading to what we called  DKNL)). The instability that continues to exist for $\Theta_e >4$ (its growth rate corresponds to the green solid line in the left panel of Fig. \ref{fig:swelling-vs-ae}) is associated to the destabilization of this extra plasma mode (while both the fast and slow modes continue to exist and remain stable). The only small difference observed in the presence of Landau damping is that the intermediate mode becomes purely imaginary in a range of values of $\Theta_e$ between 2 and 4 (see Fig. \ref{fig:minmaxint} left, dash-dotted line). We also note that the use of the $R_{42}$ Pad\'e does not change the roots associated with the slow and fast modes, but only those associated to the extra damped plasma modes (not shown). We conclude that in an angular boundary layer close to $90^\circ$, the fast mode is always stable, and the instability that continues to exist for $\Theta_e>4$ is of the same nature as the slow mode swelling instability, i.e. its growth rate tends to zero as $\alpha$ approaches $90^\circ$.

	\begin{figure}
		\centerline{
			\includegraphics[width=0.48\textwidth]{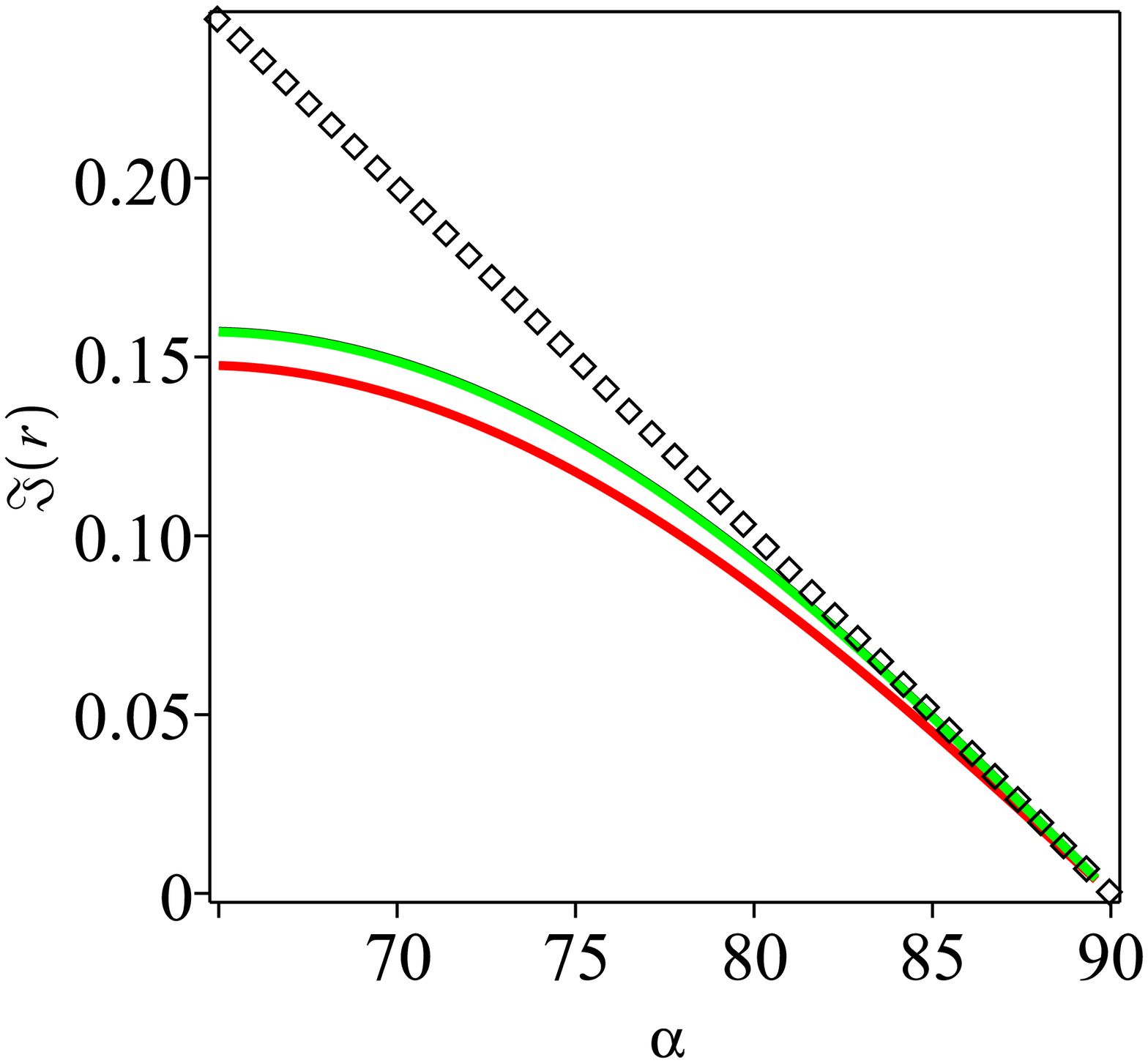}
			\includegraphics[width=0.48\textwidth]{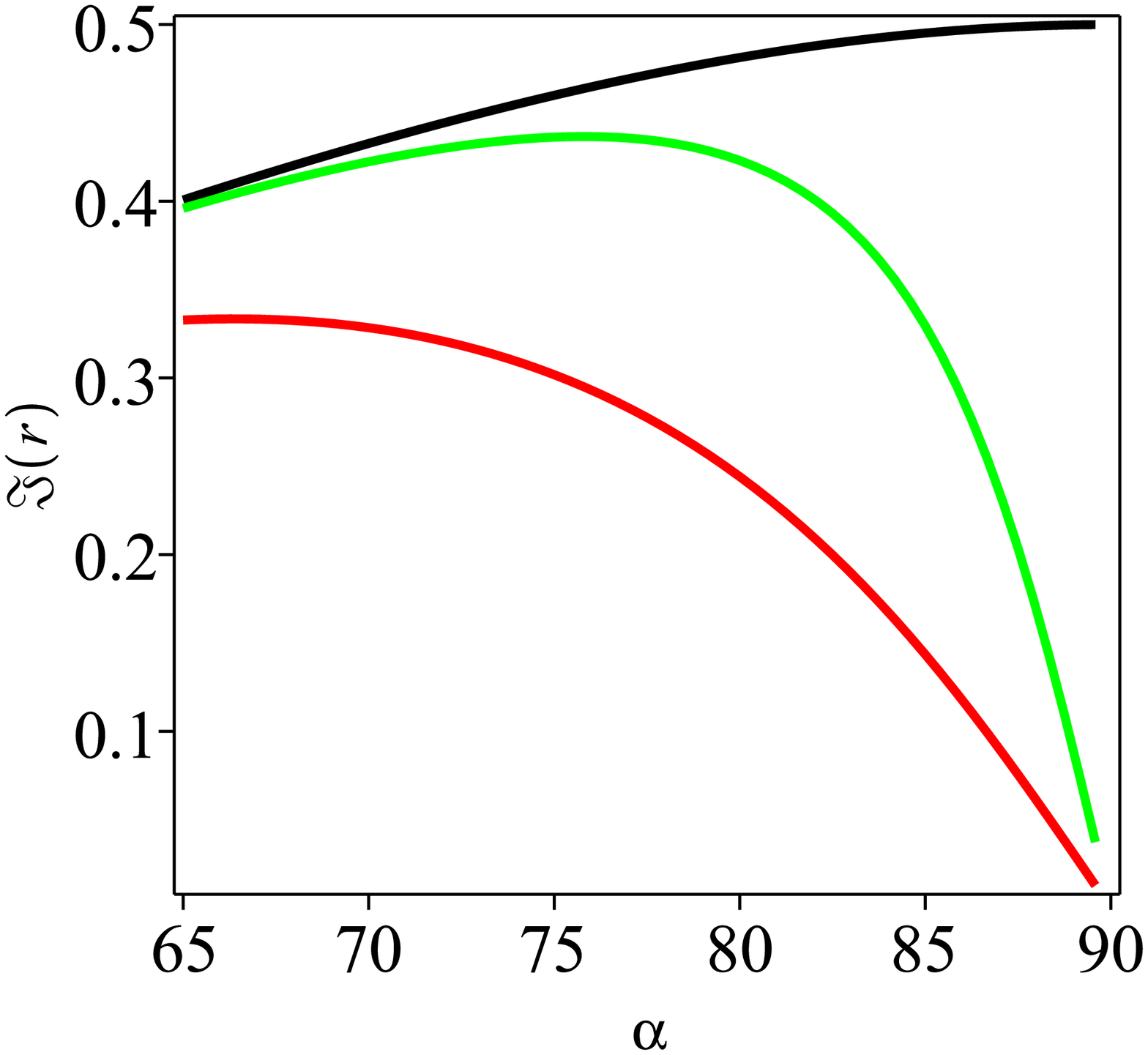}}		
		\caption{Growth rate of the unstable mode as a function of the propagation angle $\alpha$ for the BC84 (black), DK (red), DKNL (green) and  GF4 (diamond symbols) models in the case  $\Theta_e=3$ (left) and $\Theta_e=4.5$ (right).  For $\Theta_e=3$, BC84 and DKNL curves are superimposed.}
		\label{fig:swelling-vs-angle}
	\end{figure}
	
	For $\alpha=80^\circ$, the value of $\zeta_e$ is sufficiently small for the approximation $R(\zeta_e)= 1$ to be valid. In this case, the behavior of the slow mode swelling instability is similar to the previous case, the main difference affecting the fast mode. Its phase velocity for $\Theta_e<4$ now corresponds to the intermediate branch, the one with the largest real part corresponding to the extra plasma mode which, in the presence of Landau damping, is heavily damped (damping rate not shown). For $\Theta_e>4$, the real frequency of the fast mode vanishes and this mode becomes unstable, as predicted in \citet{Bas84}. The slow mode reappears on this intermediate branch with a very good match between the GF4 and DK as well as DKNL predictions.
	
	Complementary information is presented in Fig. \ref{fig:swelling-vs-angle} which displays the growth rate of the unstable mode as a function of the angle $\alpha$ for the three models described above for $\Theta_e=3$ (left) and $\Theta_e=4.5$ (right). For the case corresponding to the slow mode swelling instability of \citet{Bas84} ($\Theta_e=3$), the three models are very similar (the case without Landau damping is actually almost identical to the BC84 model so that both curves are superimposed). The GF4 model gives very similar growth rates for angles between $80^\circ$ and $90^\circ$ but its predictions deviate for smaller angles. For $\Theta_e=4.5$  the fast mode instability as predicted by BC84  displays a growth rate proportional to $k$ up to angles $\alpha=90^\circ$. This fast-mode instability is only  recovered with DKNL (and DK) at oblique angles, the deviation with BC84 starting to be significant for $\alpha > 75^\circ$. For this value of $\Theta_e$, the slow mode is always stable in the GF4 model.

	\bigskip
	 {\bf Remarks:}\\
	
	\begin{itemize}
		\item{\it Influence of warm ions}:
		If one considers the fast mode at an angle close (but not equal) to $90$ degrees, one can assume warm ions and at the same time $R(\zeta_i)= 0$. Taking also $R(\zeta_e)=1$, and using Eq. (\ref{disp}), one gets the dispersion relation (Eq. 20) obtained by \citet{Pokhotelov14} who show  that ions are stabilizing.
		
		\item{\it The case with ion temperature anisotropy:}
		With finite (but isotropic) ion temperatures, other modes are present, but the one which becomes first unstable when electron temperature anisotropy is increased is still the slow mode.
		The case where the instability is driven by ion temperature anisotropy (so-called classical mirror instability with isotropic electrons) is in contrast different since the instability originates from the extra mode associated with the finite ion temperature fluctuations and not from the slow mode which continues to exist and to be stable above the mirror threshold. An interesting point is that the mirror mode, usually thought of being non-propagating, originates from one of the damped propagating "ion temperature modes". These modes only become non-propagating for a large enough ion temperature anisotropy.
	\end{itemize}

\end{appendix}

\bibliographystyle{jpp}
\bibliography{tassisecondrevised}

\end{document}